\documentclass[preprint,12pt,square,comma,number]{elsarticle}
\usepackage{numcompress}
\usepackage{amsmath}
\usepackage{color,txfonts}
\usepackage{amssymb}
\usepackage{graphicx}
\usepackage{amssymb}
\usepackage{verbatim}
\usepackage{booktabs}
\usepackage{subfig}
\usepackage[colorlinks,
            linkcolor=blue,
            anchorcolor=blue,
            citecolor=blue]{hyperref}
\usepackage{lineno}
\usepackage{siunitx}

\begin{document}

\begin{frontmatter}

\title{A Telescope System for Charge and Position Measurement of High Energy Nuclei}

\author[1,2]{Dexing~Miao\corref{cor2}}
\author[1,12]{Zhiyu~Xiang\corref{cor2}}
\author[7]{Giovanni~Ambrosi}
\author[7]{Mattia~Barbanera}
\author[1]{Baasansuren~Batsukh}
\author[1]{Mengke~Cai}
\author[3]{Xudong~Cai}
\author[6]{Yuan-Hann~Chang}
\author[1]{Shanzhen~Chen}
\author[6]{Hsin-Yi~Chou}
\author[1]{Xingzhu~Cui}
\author[1,2]{Mingyi~Dong}
\author[7]{Matteo~Duranti}
\author[1]{Ke~Gong}
\author[1,2]{Mingjie~Feng}
\author[13]{Valerio~Formato}
\author[4,5]{Daojin~Hong}
\author[7]{Maria~Ionica}
\author[1,2]{Xiaojie~Jiang}
\author[7]{Yaozu~Jiang}
\author[1]{Liangchenglong~Jin}
\author[1,9]{Shengjie~Jin}
\author[3]{Vladimir~Koutsenko}
\author[1,10]{Tiange~Li}
\author[1,2]{Zuhao~Li}
\author[6]{Chih-Hsun~Lin}
\author[4]{Cong~Liu}
\author[4]{Pingcheng~Liu}
\author[1,2]{Xingjian~Lv}
\author[8]{Alberto~Oliva}
\author[1,2]{Ji~Peng}
\author[1]{Wenxi~Peng}
\author[1]{Rui~Qiao}
\author[1,2]{Shuqi~Sheng}
\author[7]{Gianluigi~Silvestre}
\author[1]{Congcong~Wang}
\author[1]{Feng~Wang}
\author[4]{Hongbo~Wang}
\author[5]{Zibing~Wu}
\author[4]{Suyu~Xiao}
\author[4,5]{Weiwei~Xu}
\author[1]{Sheng~Yang}
\author[1]{Xuhao~Yuan}
\author[1]{Xiyuan~Zhang}
\author[1]{Zijun~Xu\corref{cor1}}\ead{xuzj@ihep.ac.cn}
\author[1]{Jianchun~Wang\corref{cor1}}\ead{jwang@ihep.ac.cn}

\cortext[cor2]{Co-first authors.}
\cortext[cor1]{Corresponding author}

\address[1]{Institute of High Energy Physics, Beijing 100049, China}
\address[2]{University of Chinese Academy of Sciences, Beijing 100049, China}
\address[12]{Central South University, Changsha 410083, China}
\address[3]{Massachusetts Institute of Technology (MIT), Cambridge, Massachusetts 02139, USA}
\address[4]{Shandong Institute of Advanced Technology, Jinan
250100, China}
\address[5]{Shandong University (SDU), Jinan, Shandong 250100, China}
\address[6]{Institute of Physics, Academia Sinica, Nankang, Taipei, 11529, Taiwan}
\address[7]{INFN Sezione di Perugia, 06100 Perugia, Italy}
\address[8]{INFN Sezione di Bologna, 40126 Bologna, Italy}
\address[9]{Northwest Normal University, Lanzhou 730070, China}
\address[10]{Hunan University, Changsha 410082, China}
\address[13]{INFN Sezione di Roma2, 00133 Roma, Italy}

\begin{abstract}
A high-granularity telescope system with a large sensitive area and low material budget has been developed for high-energy heavy ion beam tests. The telescope consists of nine layers of silicon microstrip detectors (SSDs), whose performance was validated through a heavy ion beam test at the CERN SPS. A hybrid machine learning algorithm is proposed to address the challenges of nuclear charge measurement with SSDs. The system achieves a spatial resolution of $\mathcal{O}(1) \,$\SI{}{\micro\metre} and a charge resolution better than 0.16 charge units for nuclei from $Z = 1$ to $Z = 29$, with a sensitive area of $8 \times 8 \, \mathrm{cm}^2$. To the best of our knowledge, this represents the most precise charge and spatial resolution simultaneously achieved by a silicon telescope to date.
\end{abstract}

\begin{keyword}
Silicon Microstrip Detector, Heavy Ion Beam Test, Machine Learning, Charge Resolution, Spatial Resolution
\end{keyword}
\end{frontmatter}

\section{\label{sec:intro}Introduction}

Experimental research in nuclear and particle physics relies on the development of advanced particle detectors~\cite{LHCb:2008vvz,HIAF,Sadrozinski:2001ck,Lubelsmeyer:2011zz}. Testing with high-energy particle beams is essential to verify the correct operation and to conduct quantitative performance studies during detector development. In studies involving nuclei, heavy ion beam tests are frequently necessary. Since some heavy-ion beam test facilities provide mixed beams resulting from fragmentation~\cite{SPS,BNL,PS}, different nuclei with various charges are present simultaneously in the beam. Therefore, an independent charge identification is essential in beam tests in order to reliably determine the nuclear charge ($Z$) of each incident ion. In addition, evaluating the performance across different regions of a detector or studying the spatial resolution requires precise particle track information. 

To meet these two requirements, we have developed a telescope system based on silicon microstrip detectors (SSDs), which enables nuclear charge identification for ions with \( Z = 1 \) to $Z = 29$ and provides spatial measurements with $\mathcal{O}(1)$ \SI{}{\micro\metre} precision. The telescope has been successfully employed in beam tests for both the AMS-02~\cite{Lubelsmeyer:2011zz} Layer-0 tracker upgrade~\cite{Li:2026jxg,Miao:2025ldv} and the HERD~\cite{HERD:2014bpk} detector development.

In this paper, we present the design of the telescope, its characterization in a heavy ion beam test, a hybrid machine learning algorithm for charge measurement, and the resulting charge and spatial resolution.

\section{Telescope design with silicon microstrip detector}

\begin{figure}[h]
\centering
\subfloat[\label{fig:SSD_frame_board}]{\includegraphics[width=0.45\hsize]{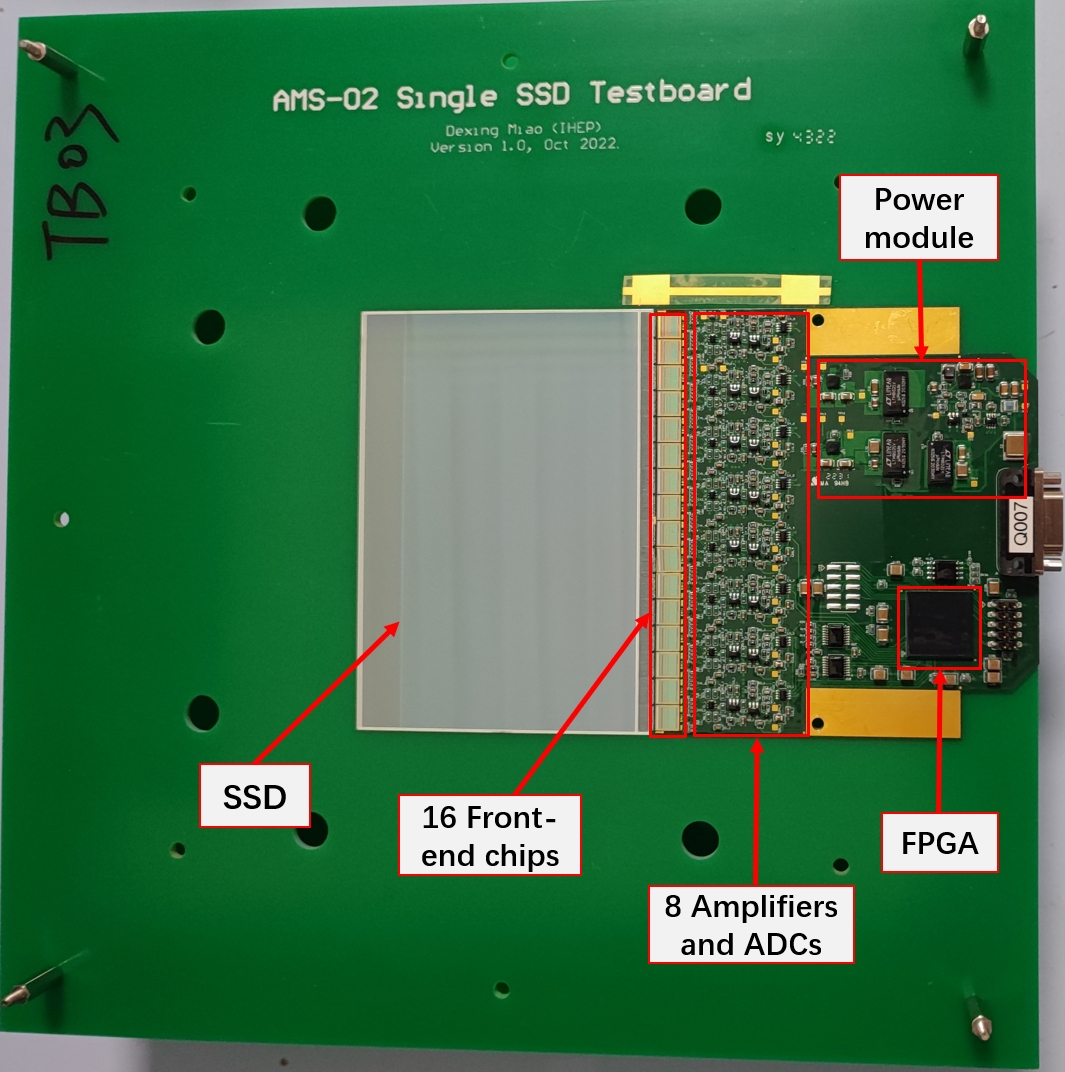}}
\hspace{0.2cm}
\subfloat[\label{fig:beam_monitor}]
{\includegraphics[width=0.35\hsize]{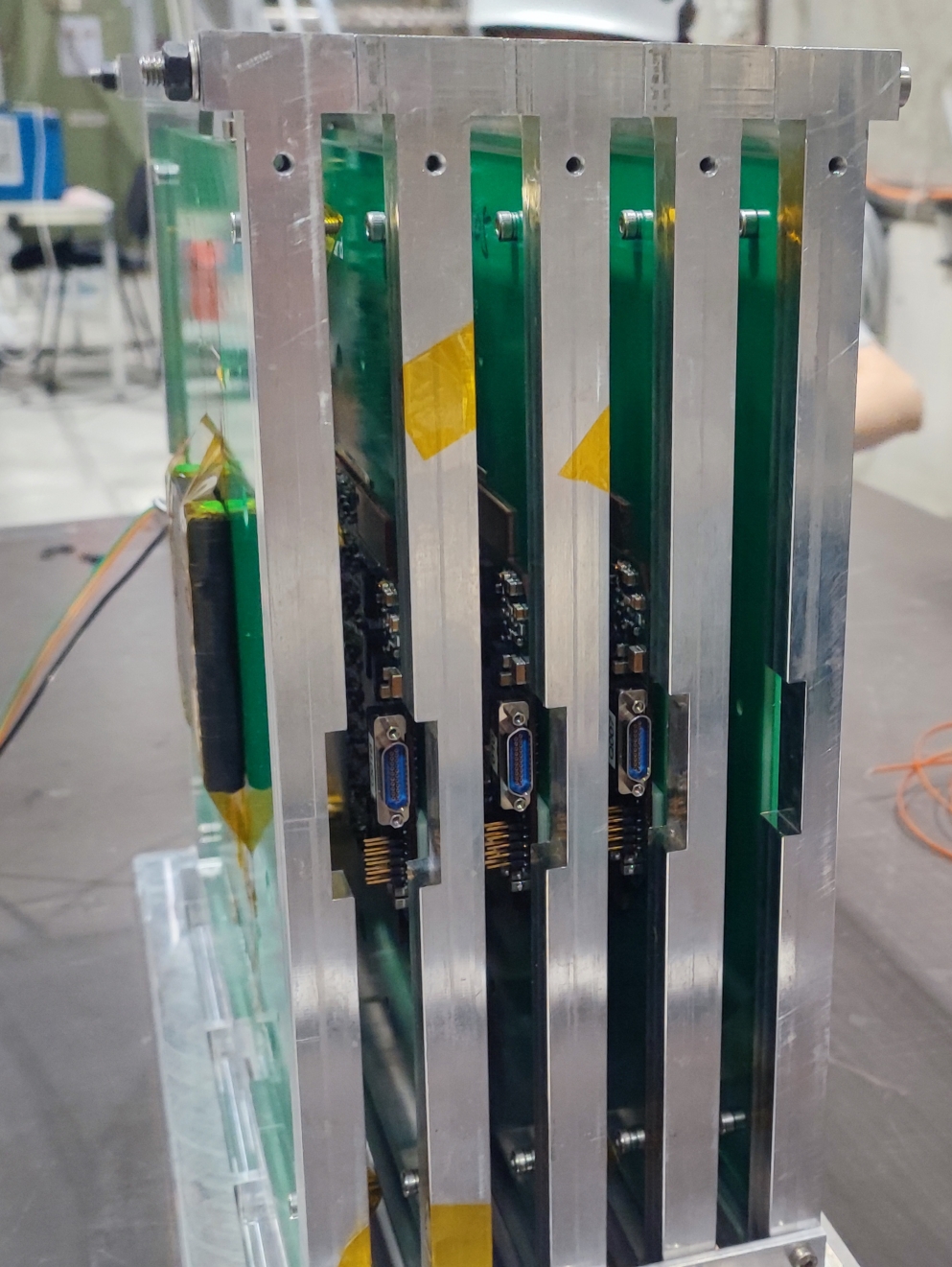}} 
\caption{(a) Photograph of a single SSD test board (SSTB), with the different components labeled. (b) Photograph of the telescope with SSTBs mounted on the aluminum support frames.}
\label{fig:subfigures}
\end{figure}

\begin{figure}[htbp] 
\centering 
\includegraphics[width=0.8\hsize]{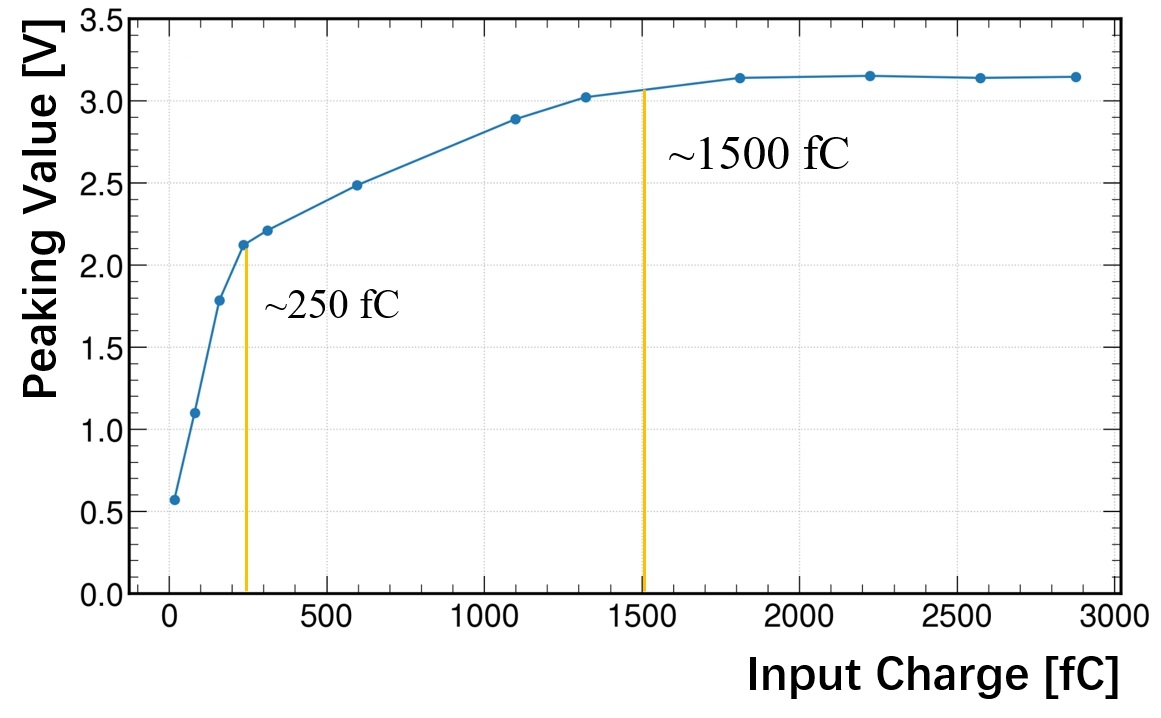}  
\caption{Response curve of the IDE1140 front-end chip obtained from a pulse injection study. The curve consists of two linear regions: the first from 0 to $250\, \mathrm{fC}$, and the second extending up to $1500\, \mathrm{fC}$.\label{va_response}}
\end{figure}

The detailed design and performance characterization of the SSD used in this work have been described in Ref.~\cite{Miao:2025ldv}. Several single SSD test boards (SSTBs) have been designed as shown in the left panel of Fig.~\ref{fig:subfigures}. The gray rectangle in the center is the SSD with an active area of $11 \times 8\, \mathrm{cm}^2$ and a thickness of \SI{320}{\micro\metre}, which has 4095 $\mathrm{p}^{+}$ strips with a pitch of \SI{27.25}{\micro\metre}. Every fourth $\mathrm{p}^{+}$ strip, starting from the second one, is AC-coupled to a readout aluminum strip, referred to as a readout strip. The three strips located between two readout strips are called floating strips. The floating strips enhance the charge-sharing effect, thereby improving both the spatial resolution and the charge resolution of the detector.

The 1024 readout aluminum strips of the SSD are wire-bonded to 16 front-end readout chips, IDE1140~\cite{IDE1140}, each containing 64 input channels. Fig.~\ref{va_response} shows the response curve of the IDE1140 chip obtained from a pulse injection study using a signal generator and an external capacitor. This curve consists of two linear regions: the first spans $0\text{--}250\, \mathrm{fC}$ with a steeper slope, and the second extends from $250\text{--}1500\, \mathrm{fC}$ with a reduced gain. A minimum ionizing particle (MIP) with unit charge ($Z = 1$) traversing \SI{320}{\micro\metre} of silicon generates approximately $4\, \mathrm{fC}$ of electron--hole pairs. Therefore, a single readout channel is roughly capable of identifying nuclei with charge up to $Z \approx \sqrt{1500/4} \approx 19$. Taking into account the charge-sharing effect among neighboring readout strips, multi-channel information can further improve the charge identification capability.

The telescope is assembled from several layers of SSTBs, as depicted in the right panel of Fig.~\ref{fig:subfigures}. Two SSTBs are mounted on each aluminum support frame, and the SSTB is designed with 90-degree rotational symmetry, allowing flexible orientation of each layer. In this work, we employed 4 vertical layers and 5 horizontal layers, which collectively enable two-dimensional track reconstruction.

With this telescope, a heavy ion beam test was carried out at the CERN Super Proton Synchrotron (SPS) in November 2023. The ion beam was produced by fragmenting a lead primary beam of $150\, \mathrm{GeV/n}$ on a beryllium target. The resulting secondary beam was selected by magnetic optics with the mass-to-charge ratio ${A}/{Z} = 2$. Since the fragmentation process at high energy preserves the projectile velocity, all selected fragments carry approximately the same momentum per nucleon as the primary beam, i.e., $\sim 150\, \mathrm{GeV/n}$.

Two plastic scintillators~\cite{Zhang:2023mti} were installed upstream and downstream of the telescope to provide trigger signals. A charge tagger detector (CT)~\cite{Adriani_2025} was placed upstream of the telescope to provide independent charge information for incident nuclei. The CT was constructed from six silicon photodiodes, each with an active area of $0.92 \times 0.92\, \mathrm{cm}^2$. The charge measurement performance of the CT with heavy ion beams is reported in Figs.~6--8 of Ref.~\cite{Adriani_2025}. We use the CT measurements for the selection of training and testing samples as described in Sec.~\ref{sec:training_set}.

\section{Hybrid machine learning algorithm for nuclei charge measurement}
\subsection{Previous methods of charge measurement with SSD}
According to the Bethe--Bloch formula~\cite{ParticleDataGroup:2024cfk}, the energy deposited by a high energy nucleus traversing a detector is proportional to $Z^2$ and the path length within the detector material. In an SSD, electron--hole pairs are generated along the particle trajectory and drift toward opposite electrodes, with the resulting charge signal collected by the readout electronics. However, the charge collection efficiency (CCE) varies with the particle's incident position relative to the readout strips. The CCE for particles incident far from the readout strips is lower than for those incident near them, with the difference reaching approximately $25\%$~\cite{ALPAT2005121}. This position dependence poses considerable challenges for particle identification (PID) among nuclei using SSDs.

An incident nucleus could generate signals on several readout channels, forming a cluster. The signal amplitude is converted by an analog-to-digital converter (ADC) to a digital value referred to as the channel value. The seed channel in a cluster has the largest value. To estimate the impact position, a variable $\eta$ is defined as the ratio of the signal amplitudes of the two highest channels in a cluster:

\begin{equation}
\eta = \frac{S_R}{S_L + S_R} ,
\end{equation}

where $S_L$ and $S_R$ are the signal values of the left and right channels among the two highest channels in the cluster, respectively. For light nuclei, hydrogen or helium for example, an $\eta$ value close to 0 (or 1) corresponds to incidence near the left (or right) readout strip, while 0.5 corresponds to incidence near the midpoint between the two readout strips. 

\begin{figure}[htbp]
\centering
\includegraphics[width=0.85\hsize]{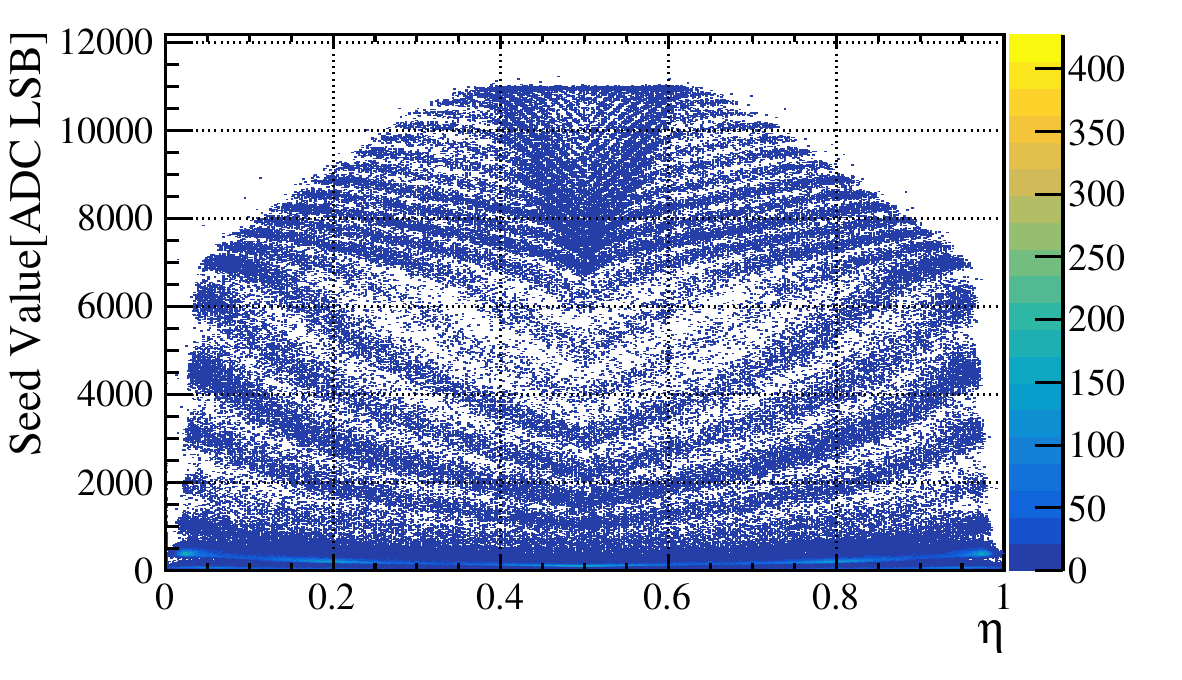}
\caption{Two-dimensional distribution of the seed channel value as a function of $\eta$. Distinct band-like structures corresponding to different nuclei are observed, each exhibiting a pronounced $\eta$ dependence.\label{eta_value:before}}
\end{figure}

As illustrated in Fig.~\ref{eta_value:before}, the two-dimensional distribution of the seed value versus $\eta$ exhibits a series of well-separated bands, each corresponding to nuclei with a specific charge $Z$. The non-horizontal structure of these bands indicates a clear dependence of the seed value on $\eta$. Furthermore, the band shapes vary among different nuclei, implying that the $\eta$ dependence differs for each charge species. Therefore, identifying different nuclei requires considering both the channel values and their correlations with $\eta$.

The AMS-02 experiment uses the ``binning $\eta$ correction'' approach~\cite{JIA2020164169,Pohl:2015pxa}: for each nucleus with known charge $Z$, the most probable values (MPVs) of the signal are calculated in each impact position bin. For an incident nucleus, the signal value is corrected according to the functional relationship between the MPV and $\eta$. However, this method is not well-suited to our case for two main reasons. First, it requires a large sample size, because each $\eta$ bin must contain enough events to ensure a reliable determination of the MPV. Second, it requires additional detector information, as the $\eta$ correction for each event depends on prior knowledge of its charge $Z$. Given that the acceptance of the CT is much smaller than that of our telescope, most events do not have such prior charge information.

The DAMPE~\cite{DAMPE:2017cev} tracker team developed a charge reconstruction algorithm called the ``Charge Sharing Algorithm''~\cite{QIAO201924}. This algorithm models the SSD and electronics as a capacitive-resistive network. Based on the linear charge-sharing assumption, it reconstructs the raw deposited energy by utilizing the signal amplitudes of the readout channels within a cluster, thereby enabling particle identification for different nuclei. However, due to the complex capacitive coupling introduced by the three floating strips in our sensor design, as well as the nonlinear effects in the front-end electronics, this method is not directly applicable to our detector.

These considerations motivate the use of a data-driven method for charge measurement. Rather than relying on explicit $\eta$ corrections or linear charge-sharing assumptions, such a method can directly exploit the multidimensional correlations among the channel values within a cluster. In this work, we adopt a machine-learning-based approach to reconstruct the nuclear charge from the cluster signals. In particular, a Boosted Decision Tree (BDT)~\cite{drucker1997improving} algorithm is employed to learn the relationship between the readout pattern and the particle charge.

\subsection{BDT approach and training dataset preparation \label{sec:training_set}}

To implement the data-driven charge reconstruction method described above, we employ a BDT regressor as the core algorithm. Tree-based models are particularly well-suited to this task because they can effectively capture complex nonlinear relationships among multiple input variables. Another advantage is that, compared with neural networks, decision-tree-based models can naturally handle input features spanning several orders of magnitude without requiring explicit normalization, since they rely on threshold-based data partitioning rather than gradient-based continuous optimization. The BDT is trained as a regressor with continuous charge labels rather than a classifier with integer labels, so that the model learns the smooth mapping from signal amplitudes to charge, preserving ordinal structure and distance information among nuclei. The training data are constructed from beam test data rather than Monte Carlo simulation, thereby avoiding systematic biases arising from the difficulty of accurately simulating the complex signal response of heavy nuclei in SSDs. Once trained, the BDT directly provides charge information for all events without the need for CT.

\begin{figure}[htbp]
\centering
\subfloat[\label{dbscan_c_before}]{\includegraphics[width=0.45\hsize]{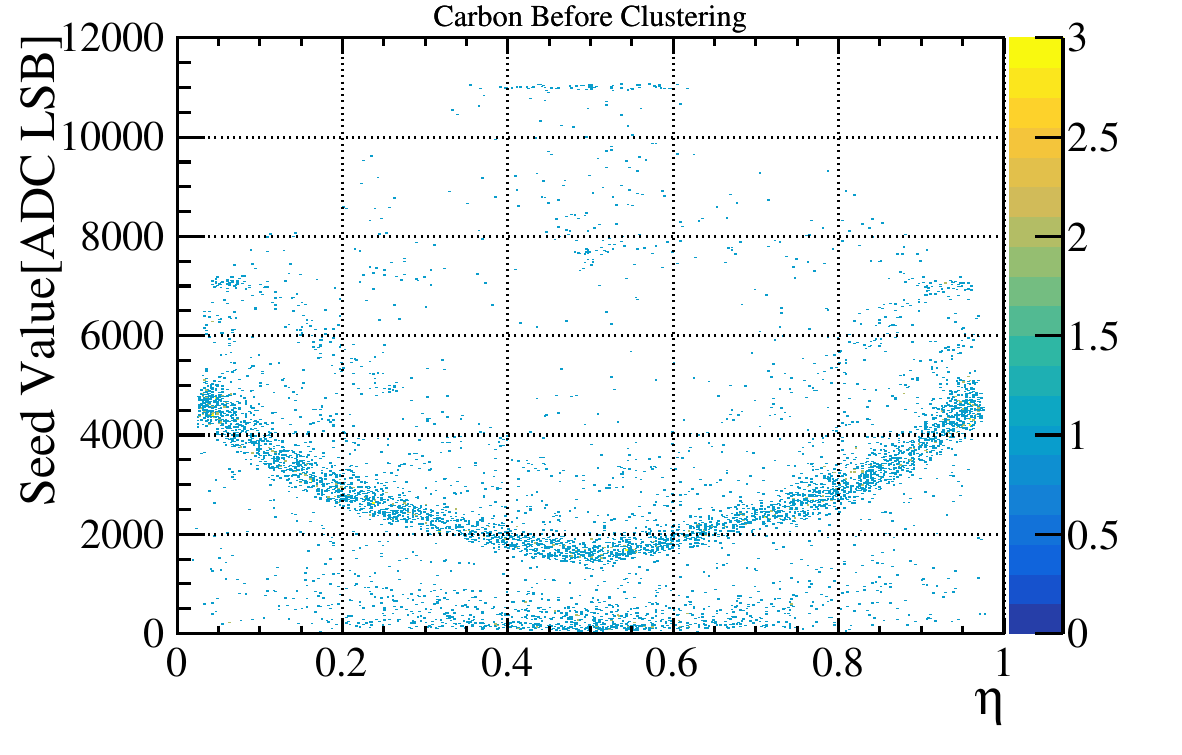}}
\hspace{0.3cm}
\subfloat[\label{dbscan_c_after}]{\includegraphics[width=0.45\hsize]{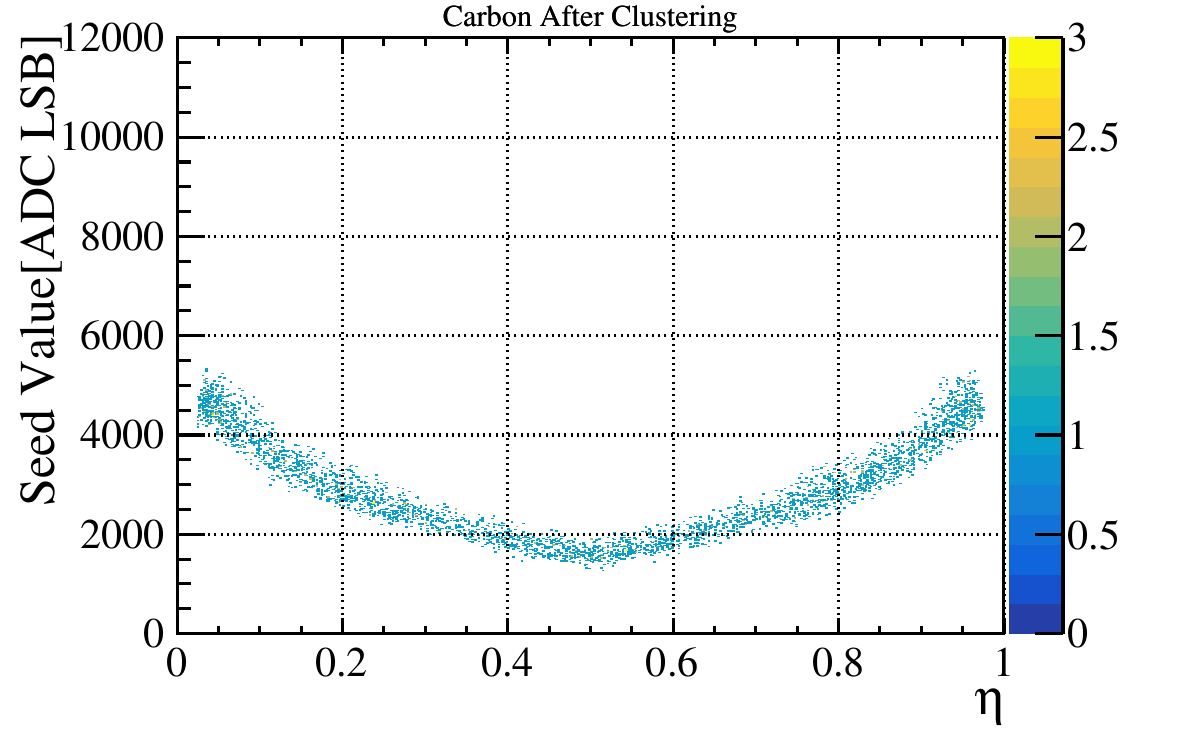}}
\\
\subfloat[\label{dbscan_s_before}]{\includegraphics[width=0.45\hsize]{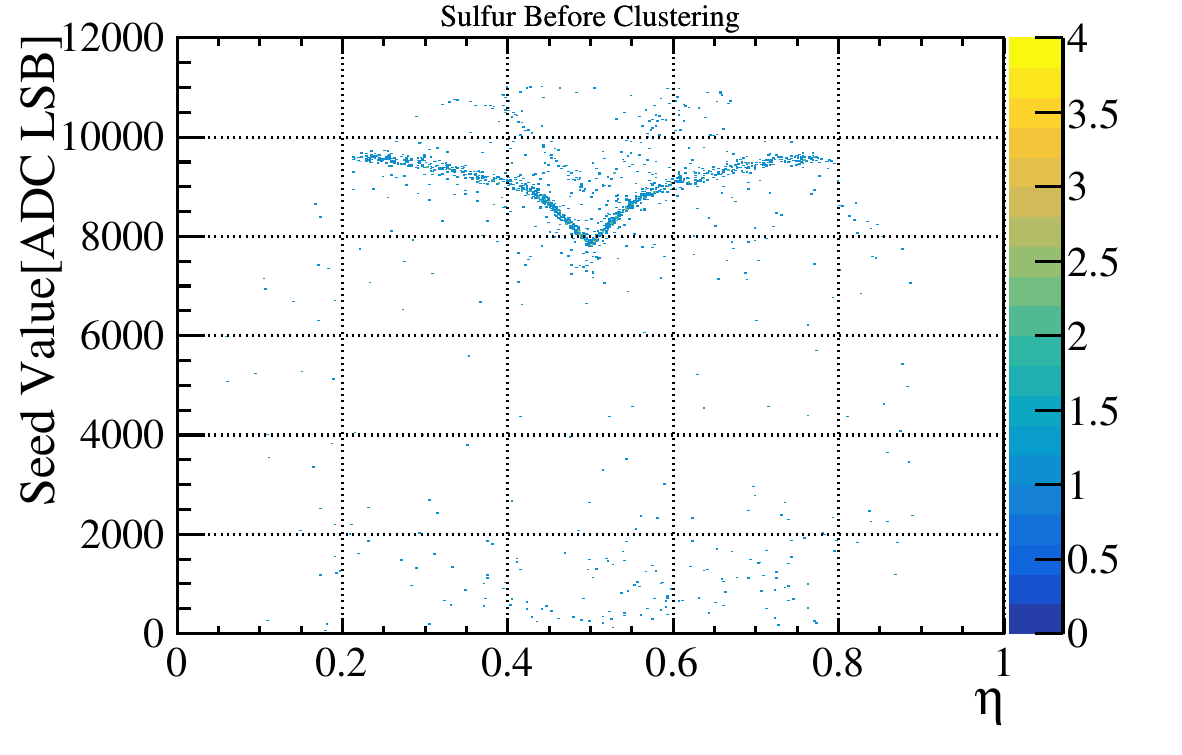}}
\hspace{0.3cm}
\subfloat[\label{dbscan_s_after}]{\includegraphics[width=0.45\hsize]{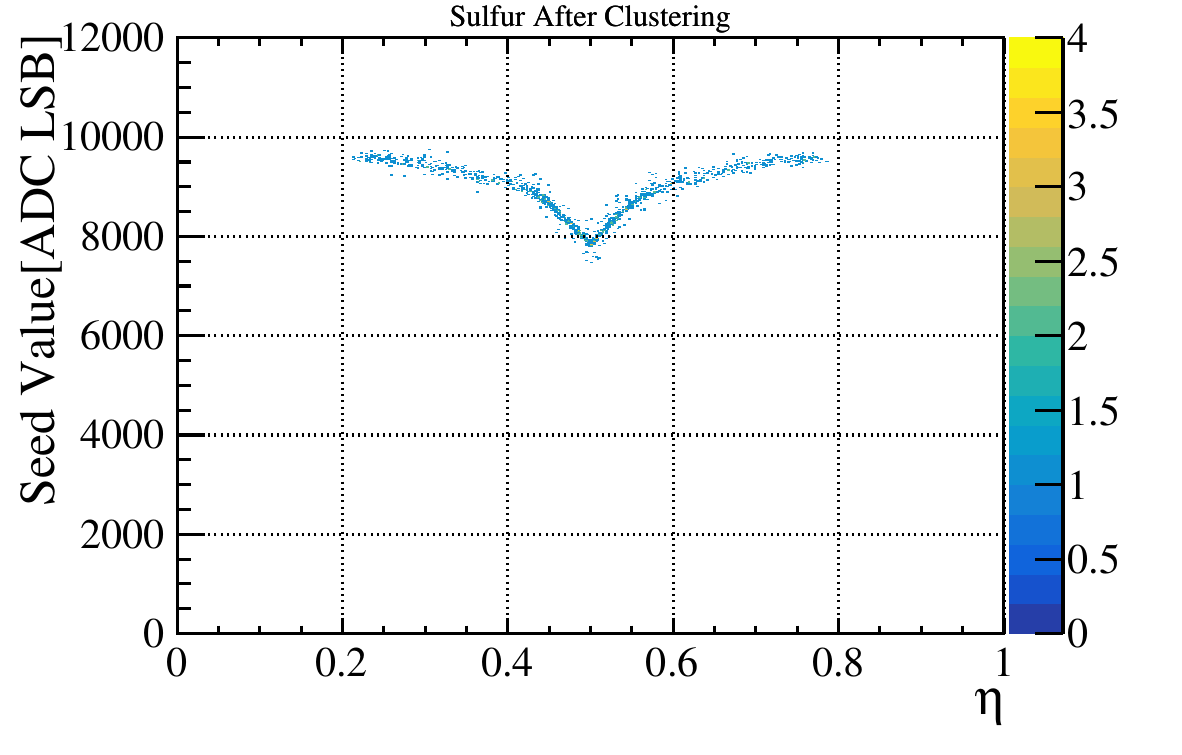}}
\caption{Seed value vs $\eta$ distribution for carbon and sulfur events, selected by charge tagger. (a) Carbon before DBSCAN; (b) Carbon after DBSCAN; (c) Sulfur before DBSCAN; (d) Sulfur after DBSCAN.} \label{eta_value_clustering}
\end{figure}

As the BDT is a supervised learning method, the first step is to prepare labeled training data. According to Ref.~\cite{Adriani_2025}, the charge resolution of the CT is about 0.25 charge units for most nuclei. For nuclei with charge $Z$, events with CT charge in the range $[Z - 0.4, Z + 0.4]$ are selected. Fig.~\ref{eta_value_clustering}\,(a) shows the selected carbon events based on CT charges. A distinct band confirms that most selected events belong to the target nucleus. However, outlier points and faint structures indicate incorrect CT assignments, possibly due to the smaller geometric acceptance of the CT or to nuclear fragmentation occurring downstream of the CT. To reject these outliers, we employ a clustering algorithm, Density-Based Spatial Clustering of Applications with Noise (DBSCAN)~\cite{10.1145/3068335}, that groups data points based on density while identifying outliers as noise. The procedure is applied to all nuclei with $Z = 1$ to $Z = 30$ after CT selection, obtaining a highly pure sample for each nucleus. Fig.~\ref{eta_value_clustering} shows carbon and sulfur events before and after DBSCAN clustering.

\subsection{Charge label construction}

The CT charge measurement yields a continuous value reflecting the CT's own noise and statistical fluctuations, with no direct relation to the SSD readout response. Directly using CT values as regression labels would train the BDT to fit CT fluctuations rather than the SSD's intrinsic charge measurements. Instead, we construct continuous charge labels from the SSD signal data itself through Support Vector Regression (SVR)~\cite{awad2015support} fitting and two-dimensional interpolation, ensuring that the labels faithfully reflect the detector's own response characteristics.

\begin{figure}[htbp]
\centering
\subfloat[\label{value12}]{\includegraphics[width=0.45\hsize]{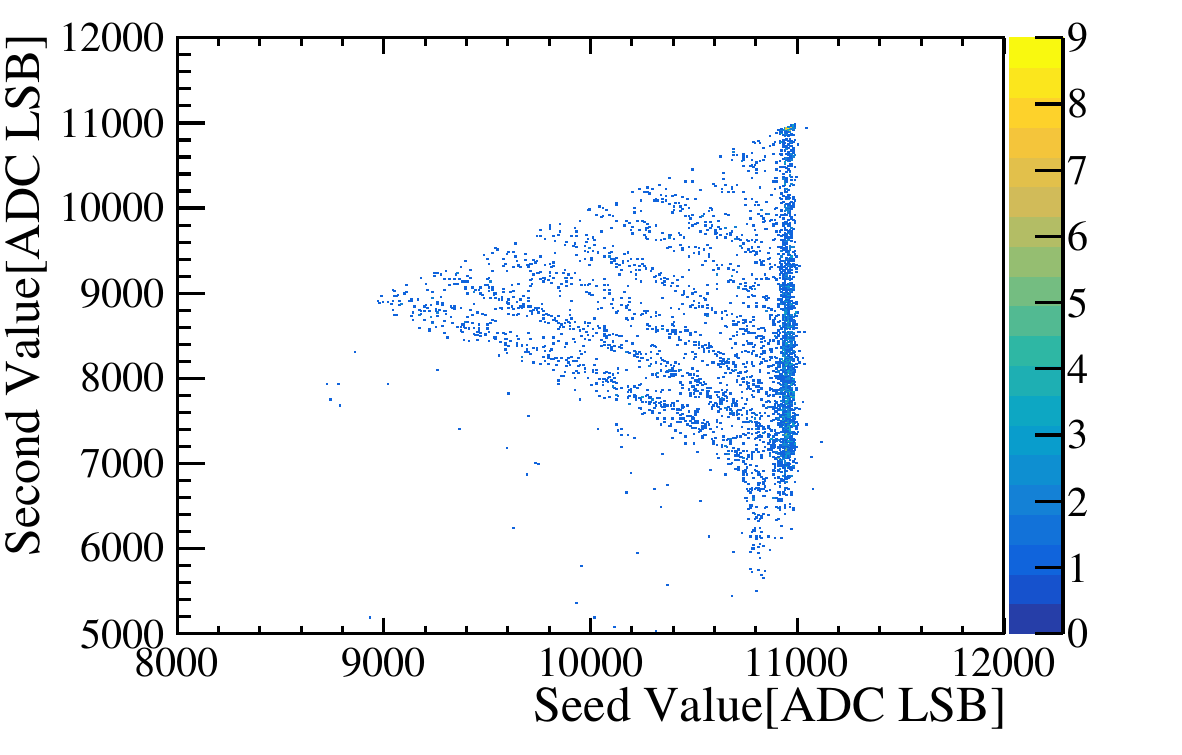}}
\hspace{0.3cm}
\subfloat[\label{value23}]{\includegraphics[width=0.45\hsize]{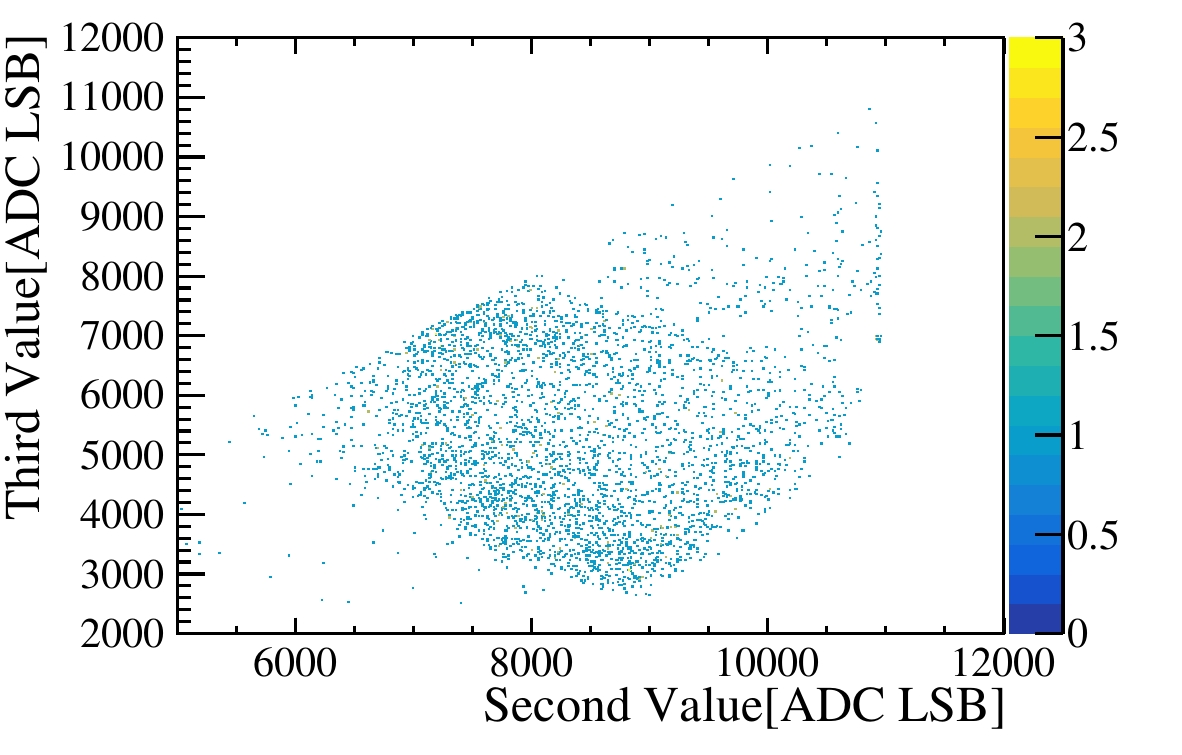}}
\caption{Correlation distributions among channel amplitudes within a cluster for $Z \geq 20$ events. (a) Second-largest versus seed channel value, showing a vertical saturation structure near seed value $\approx 1.1 \times 10^4$\,ADC\,LSB. (b) Third-largest versus second-largest channel value, retaining residual charge discrimination capability.
\label{channel_corr}}
\end{figure}

For heavy nuclei, the signal amplitude of the seed strip may saturate the dynamic range of the readout electronics. For events without saturation, the charge label is reconstructed using the signals of the two strips with the largest amplitudes in the cluster. These two channels are most closely related to the particle charge because they collect the majority of the deposited charge. For events where the seed strip is saturated, the signals of the second and the third largest strips are instead used to perform the interpolation. 

\begin{figure}[htbp]
\centering
\subfloat[\label{svr_fit_6}]{\includegraphics[width=0.45\hsize]{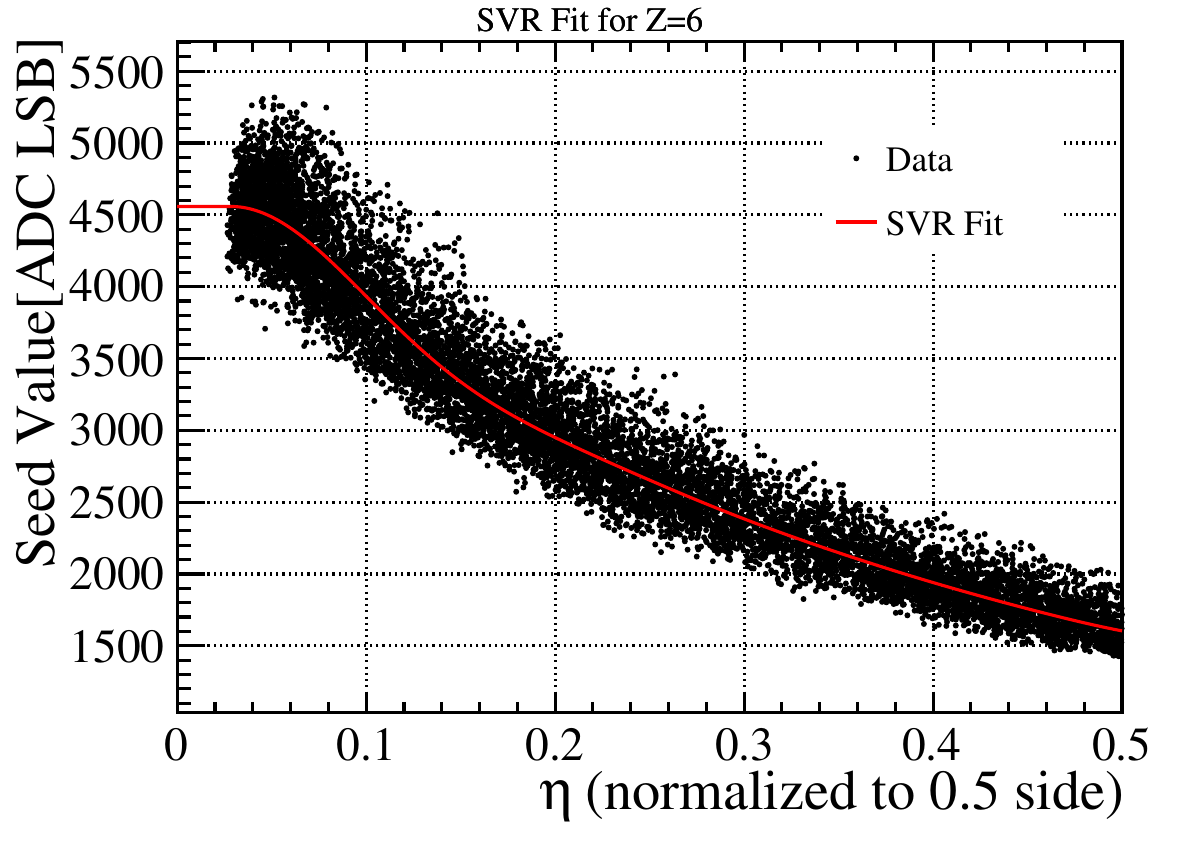}}
\hspace{0.3cm}
\subfloat[\label{svr_fit_12}]{\includegraphics[width=0.45\hsize]{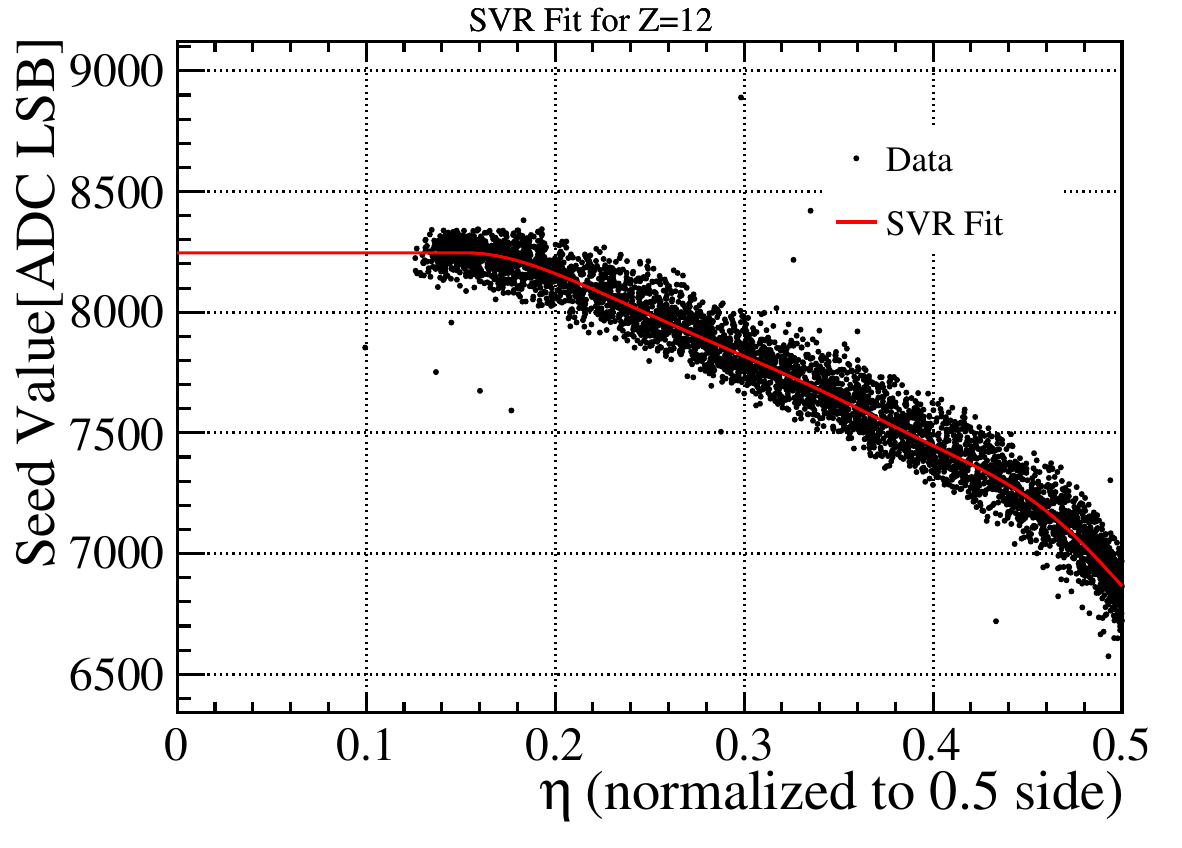}}
\caption{Seed channel value as a function of $\eta$ (normalized to the $[0,\, 0.5]$ side) together with SVR fit results. \textbf{Black dots}: data; \textbf{\textcolor{red}{Red line}}: SVR fit curve. (a) Carbon events; (b) Magnesium events.
\label{SVR_fit_seed}}
\end{figure}

Fig.~\ref{channel_corr}\,(a) shows the correlation between the second-largest and seed channel values for $Z \geq 20$ events. A pronounced vertical band is observed near a seed value of $\sim 1.1\times10^4$ ADC LSB. This structure corresponds to the saturation of the seed channel. Saturated events begin to appear from $Z \gtrsim 21$, and their fraction increases with increasing charge $Z$. For $Z \gtrsim 26$, almost all events exhibit saturation in the seed channel. In contrast, no significant saturation structure is observed in the correlation between the second- and third-largest channels shown in Fig.~\ref{channel_corr}\,(b). So, for each nucleus with $Z \leq 26$ and whose seed channel is not saturated, we use SVR to fit the relationship between the seed channel value and $\eta$ using the purified samples. Since the $\eta$ distribution is symmetric about the strip center ($\eta = 0.5$), all events are mapped to $\eta \in [0, 0.5]$ before fitting, effectively doubling the statistics and reducing the fit complexity. As illustrated in Fig.~\ref{SVR_fit_seed}, the SVR curves capture the overall trend of the seed value as a function of $\eta$ for each nucleus. Near $\eta = 0$, a data gap appears for heavy nuclei because intense charge sharing prevents $\eta$ from reaching extreme values; in this region, the fit is extrapolated with a constant value.

\begin{figure}[htbp]
\centering
\subfloat[\label{svr_fit_z22}]{\includegraphics[width=0.45\hsize]{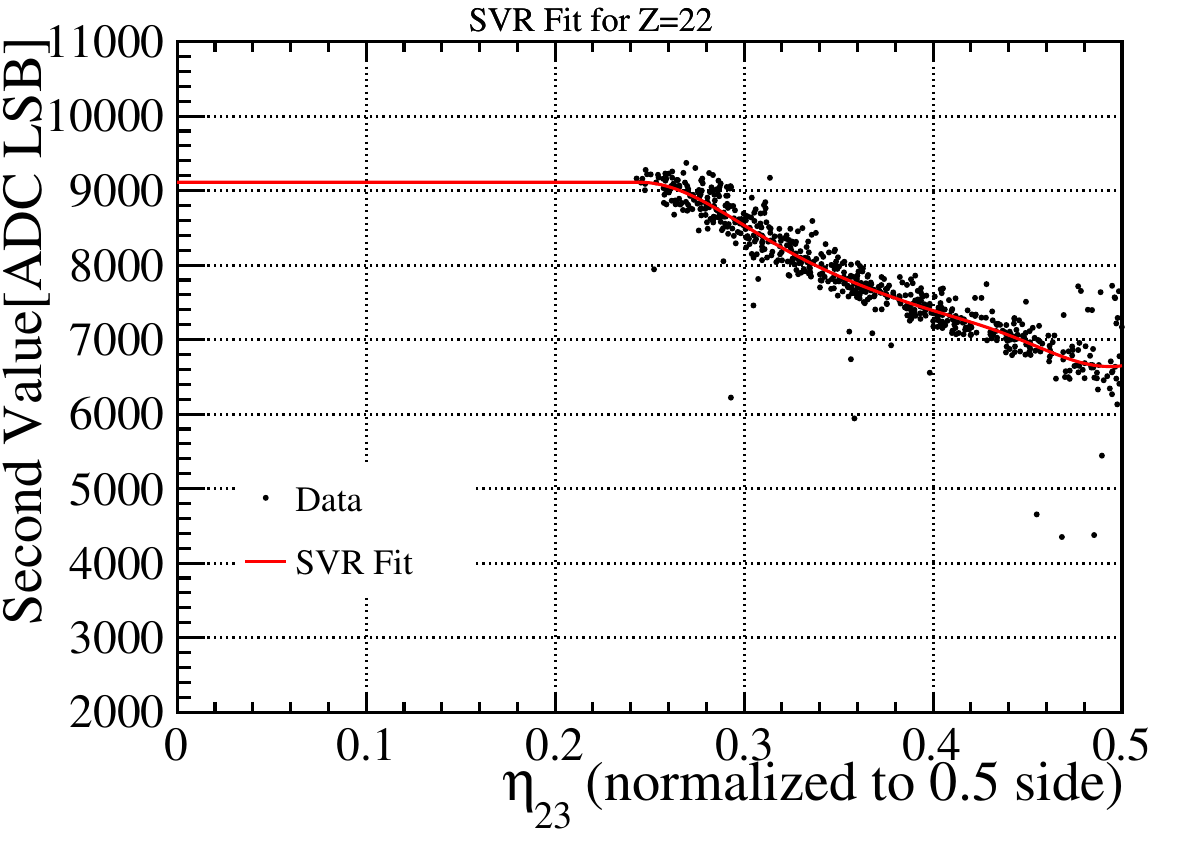}}
\hspace{0.3cm}
\subfloat[\label{svr_fit_z28}]{\includegraphics[width=0.45\hsize]{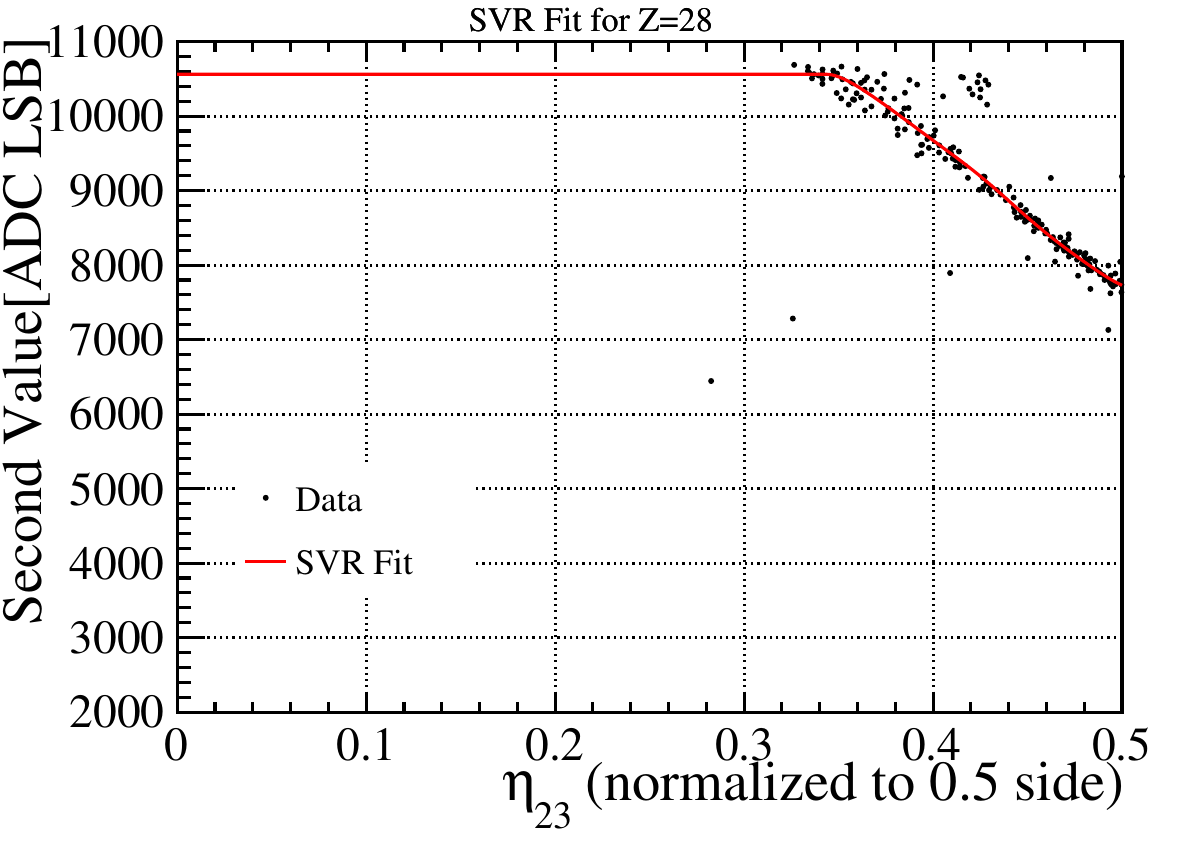}}
\caption{Second-largest channel value as a function of $\eta_{23}$ (normalized to the $[0,\, 0.5]$ side) together with SVR fit results. (a) $Z = 22$ events; (b) $Z = 28$ events.
\label{SVR_fit_23}}
\end{figure}

For events in which the seed channel is saturated, the seed value no longer reflects the information of the incident charge. To address this, we introduce a new variable analogous to $\eta$:
\begin{equation}
\eta_{23} = \frac{S_3}{S_2 + S_3},
\end{equation}
where $S_2$ and $S_3$ are the second-largest and third-largest channel values, respectively. Like $\eta$, the variable $\eta_{23}$ reflects the charge sharing ratio but is independent of the saturated seed channel. As illustrated in Fig.~\ref{channel_corr}\,(b), the second- and third-largest channels retain charge discrimination capability in the high-$Z$ region, although with reduced sensitivity compared to the seed channel, as the band structures are no longer clearly visible by eye. SVR fitting is then performed on the second-largest channel value as a function of $\eta_{23}$ for nuclei with $15 \leq Z \leq 30$, as illustrated in Fig.~\ref{SVR_fit_23}.

\begin{figure}[htbp]
\centering
\subfloat[\label{}]{\includegraphics[width=0.45\hsize]{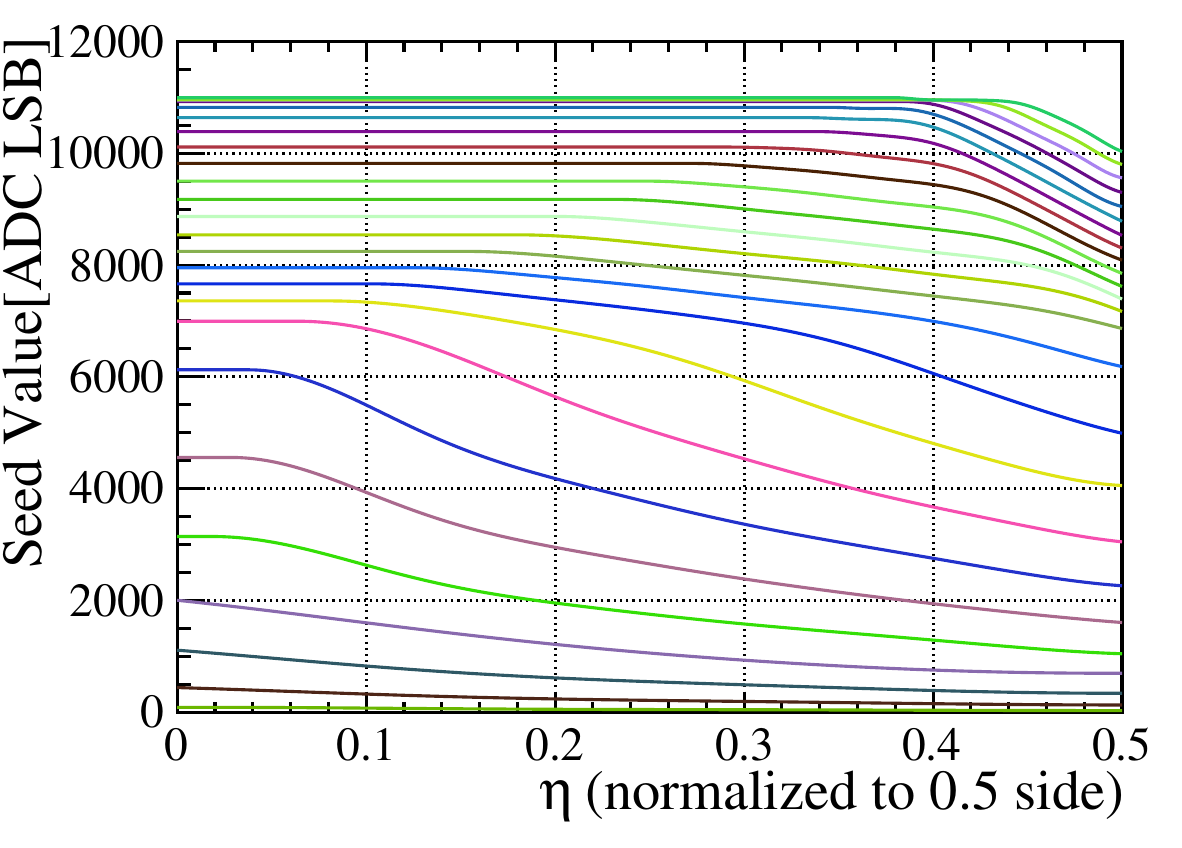}}
\hspace{0.3cm}
\subfloat[\label{}]
{\includegraphics[width=0.45\hsize]{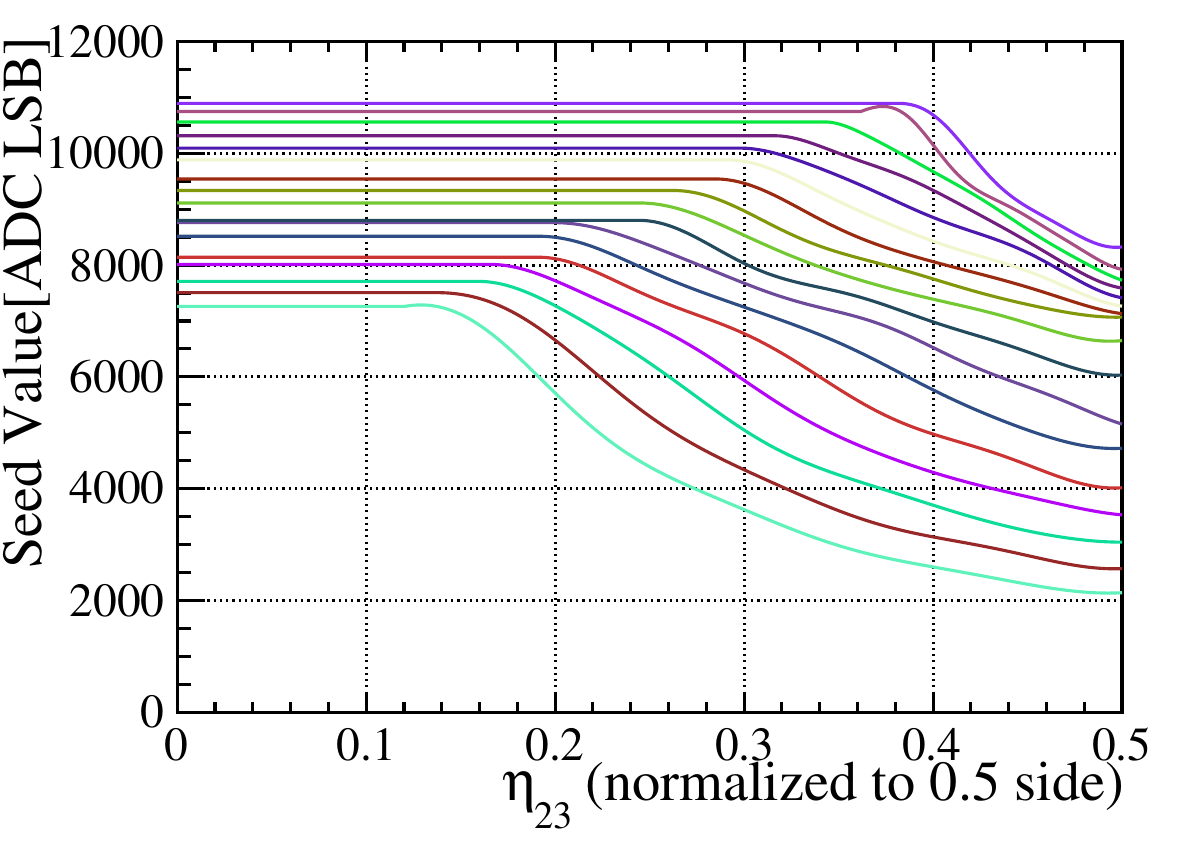}}
\caption{SVR fitting results for the channel–$\eta$ relationships of different nuclei. (a) Fit12: SVR fits of the seed channel value as a function of $\eta$ for nuclei with different charges. (b) Fit23: SVR fits of the second-largest channel value as a function of $\eta_{23}$. The curves describe the position-dependent response of the detector and provide the basis for constructing charge labels in both the non-saturated and saturated regions.\label{SVR_fit_all}}
\end{figure}

Figure~\ref{SVR_fit_all} shows the SVR fitting results for both the seed--second (Fit12) and second--third (Fit23) channel combinations. Each fitted curve corresponds to a nucleus with an integer charge $Z$. For an arbitrary event, its charge label can be obtained by interpolating among these SVR curves according to its measured $\eta$ and channel value.

\begin{figure}[htbp]
\centering
\includegraphics[width=0.55\hsize]{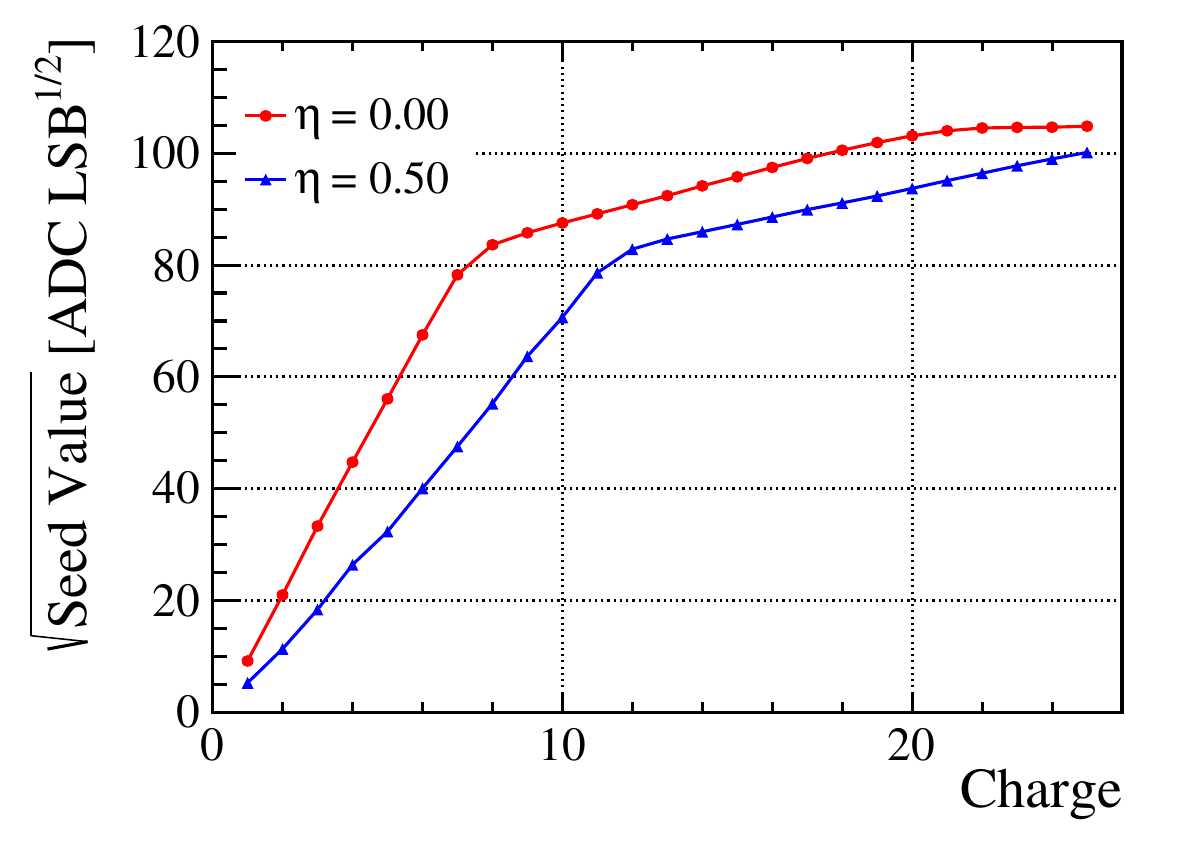}
\caption{$\sqrt{\mathrm{MPV}}$ of the seed signal amplitude at $\eta = 0$ and $\eta = 0.5$ as a function of charge number $Z$, showing a two-segment linear behavior consistent with the IDE1140 chip response.\label{mp_vs_z}}
\end{figure}

To validate the interpolation scheme, Fig.~\ref{mp_vs_z} shows $\sqrt{\mathrm{MPV}}$ of the seed signal at $\eta = 0$ and $\eta = 0.5$ as a function of $Z$. Two approximately linear regions are observed, consistent with the Bethe--Bloch relation ($\mathrm{d}E/\mathrm{d}x \propto Z^2$) together with the dual-range response of the IDE1140 chip (Fig.~\ref{va_response}). As a result, $\sqrt{\mathrm{MPV}}$ exhibits an approximately linear dependence on $Z$, which provides a natural and stable basis for charge interpolation.

For each event, a continuous charge label is obtained by performing a two-dimensional interpolation among the SVR curves corresponding to adjacent integer charges. The interpolation is carried out in the $(\eta,\, \sqrt{\text{seed value}})$ space. For events in which the seed channel is saturated, the interpolation is instead performed in the $(\eta_{23},\, \sqrt{\text{second value}})$ space. The seed--$\eta$ interpolation is used preferentially, while the second--$\eta_{23}$ interpolation provides a complementary estimate in the saturated region.

\begin{figure}[htbp]
\centering
\subfloat[\label{}]{\includegraphics[width=0.45\hsize]{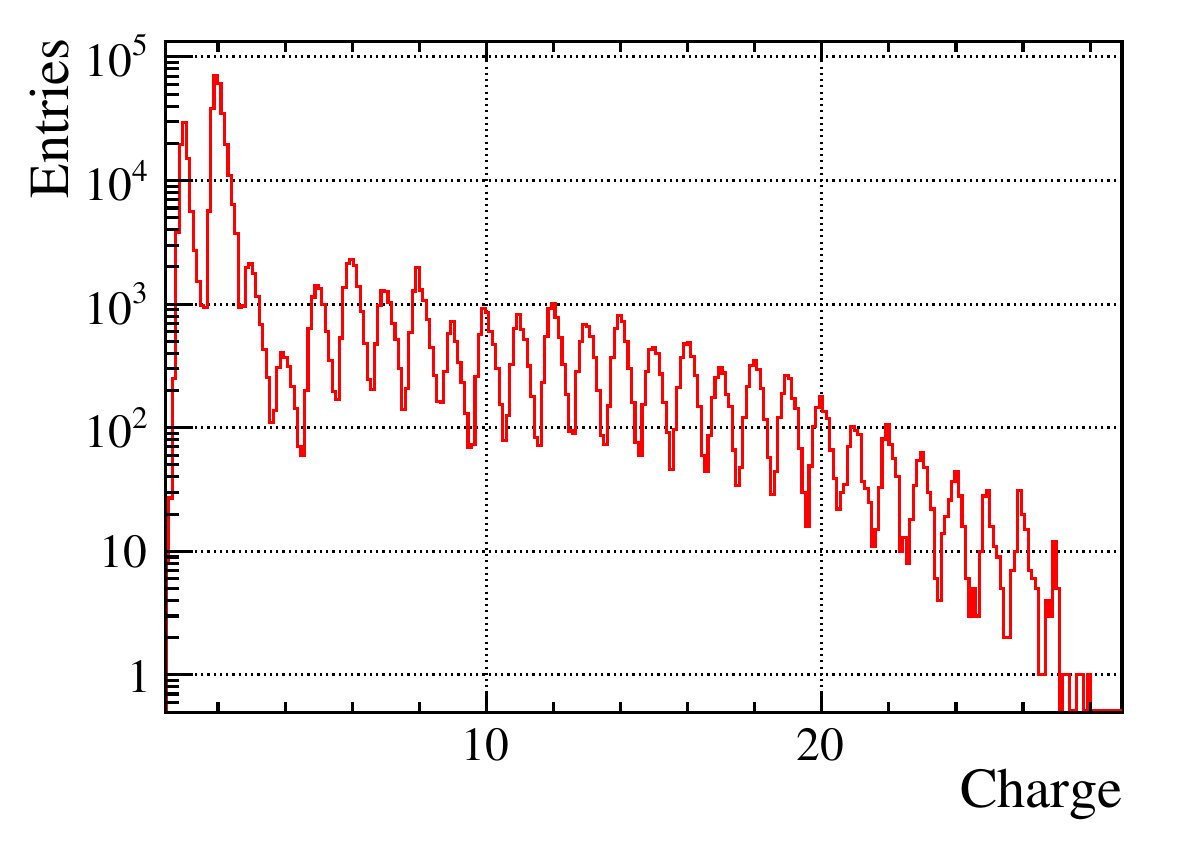}}
\subfloat[\label{}]{\includegraphics[width=0.45\hsize]{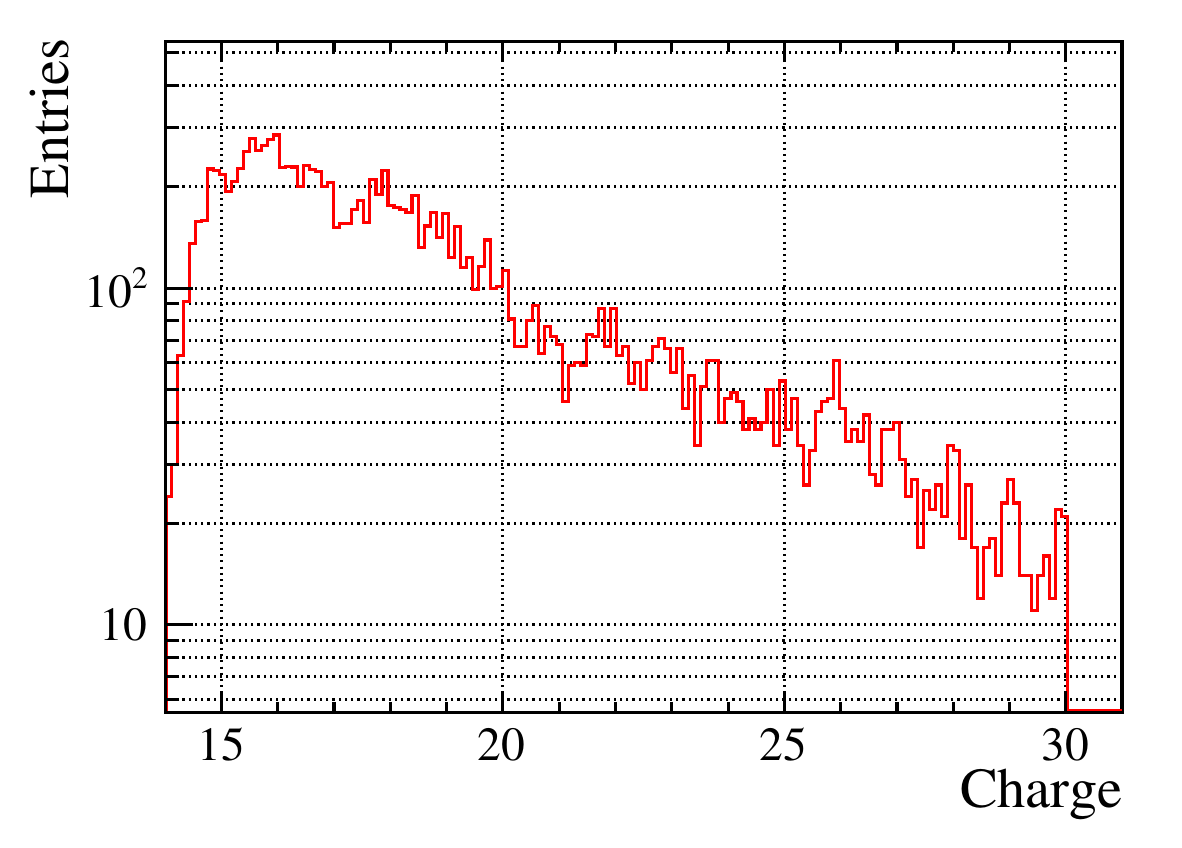}}
\caption{Continuous charge label distributions from two-dimensional interpolation. (a) Labels from seed--$\eta$ interpolation, with clear peaks up to $Z < 27$. (b) Labels from second--$\eta_{23}$ interpolation for high-$Z$ events ($15 \leq Z \leq 30$).\label{interp}}
\end{figure}

\begin{figure}[htbp]
\centering
\includegraphics[width=0.8\hsize]{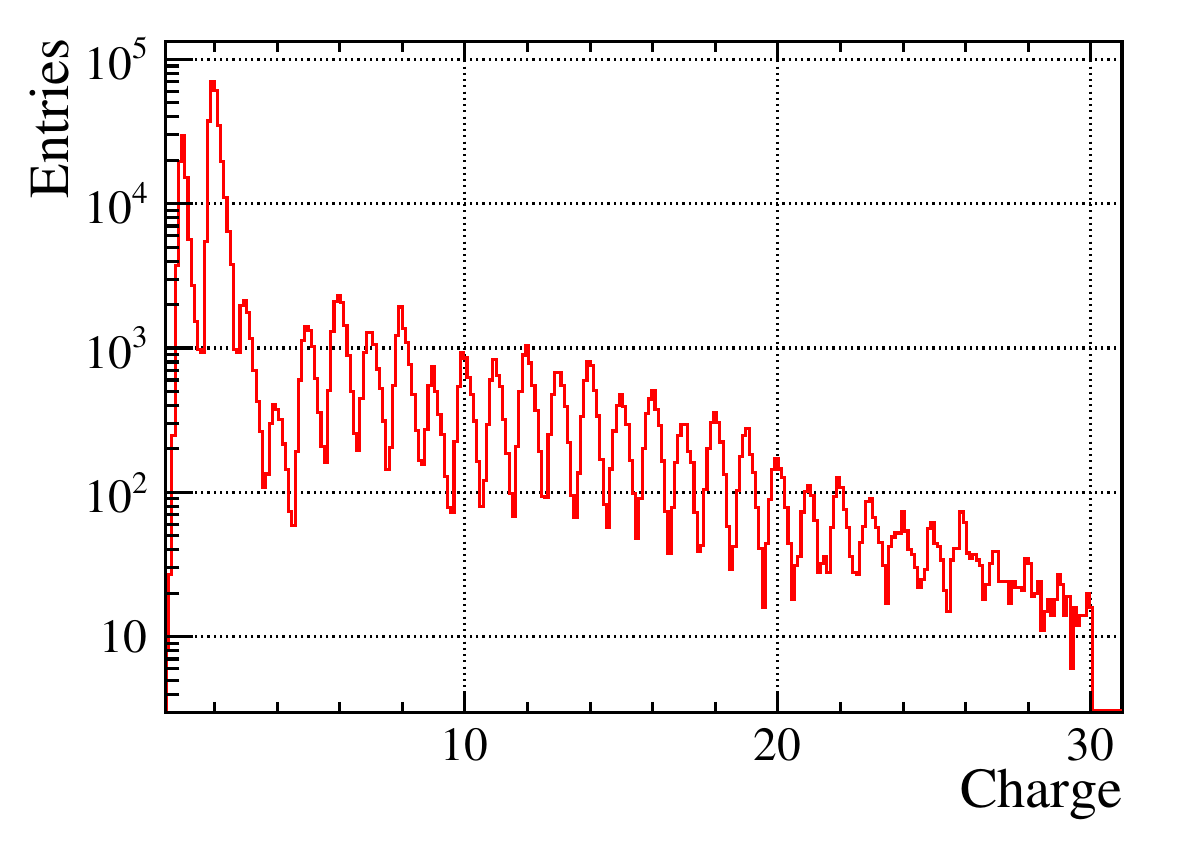}
\caption{Final charge label distribution for BDT training, combining seed--$\eta$ and second--$\eta_{23}$ interpolation results.\label{z_dist_train}}
\end{figure}

Fig.~\ref{interp} (a) shows the charge label distribution from seed--$\eta$ interpolation, with clear peaks for each integer charge up to $Z \approx 27$; beyond this, seed saturation prevents label assignment. Fig.~\ref{interp} (b) shows the second--$\eta_{23}$ result, providing labels for $Z \geq 15$ with broader peaks due to lower sensitivity of secondary channels. The final training labels are constructed by combining both sources. Since the seed--$\eta$ (Fit12) interpolation provides better charge resolution, it is used preferentially for events where the seed channel is not saturated. Only for events with seed saturation, the second--$\eta_{23}$ (Fit23) interpolation is applied instead. As a result, the interpolated charge distributions exhibit a sharper peak at lower charges, where Fit12 dominates, and become progressively broader at higher charges, where the labels are mainly derived from Fit23, as shown in Fig.~\ref{z_dist_train}.

\subsection{BDT training and testing}

\begin{figure}[htbp]
\centering
\includegraphics[width=0.8\hsize]{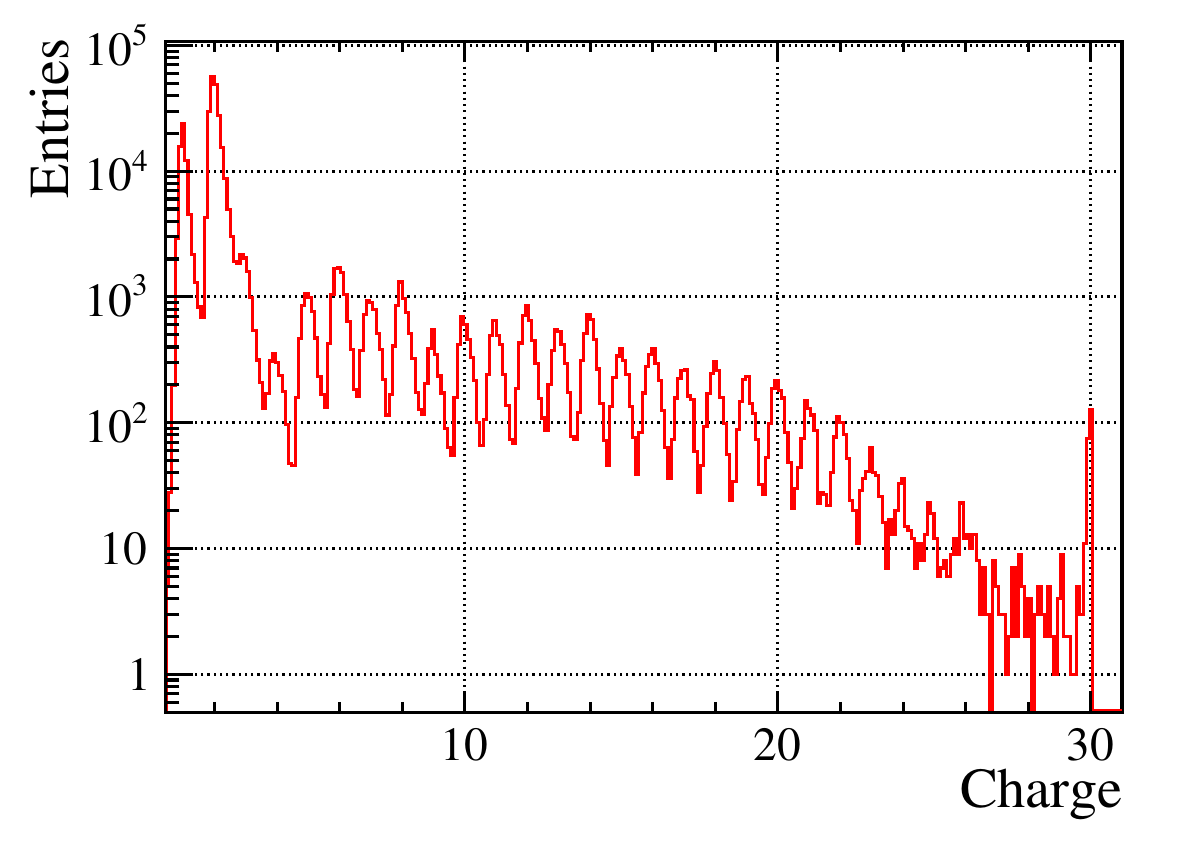}
\caption{Distribution of charge values from BDT on the testing dataset.\label{z_dist_test}}
\end{figure}

For each cluster, the three highest channel values are used as BDT input features, balancing information utilization and computational efficiency. Our telescope consists of 9 detector layers. Due to variations in SSDs and electronics, a consistently structured BDT was trained separately for each layer using $400\,\mathrm{k}$ events each for training and testing. The resulting charge distribution on the test dataset is shown in Fig.~\ref{z_dist_test}, with clear peaks at each integer charge. A pronounced peak appears at $Z=30$ because all events with true charge $\geq 30$ in the test dataset were tagged as $Z=30$. We do not investigate these events in the present work, and focus only on the charge range $Z=1\text{--}29$.

\begin{figure}[htbp]
\centering
\subfloat[\label{bdt_color_12}]{\includegraphics[width=0.45\hsize]{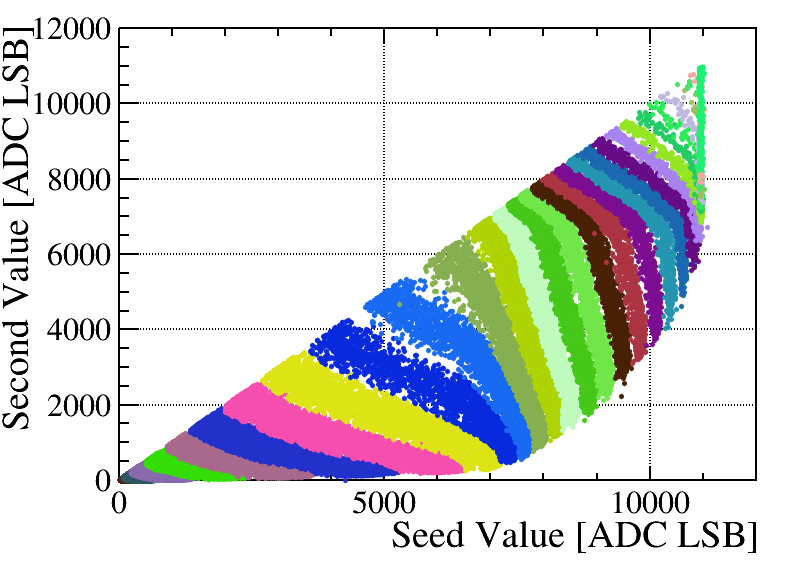}}
\hspace{0.3cm}
\subfloat[\label{bdt_color_23}]{\includegraphics[width=0.45\hsize]{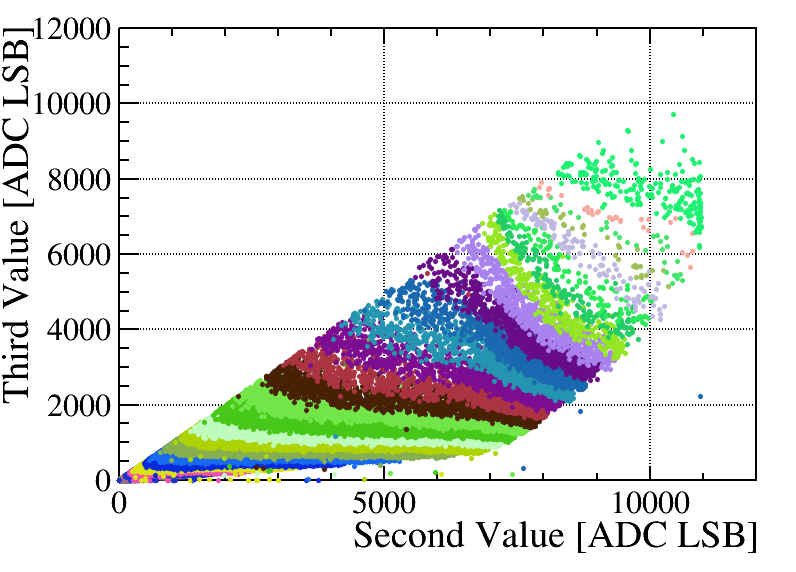}}
\caption{Channel amplitude distributions colored by BDT-predicted integer charge. (a) Seed versus second-largest channel value. (b) Second-largest versus third-largest channel value.\label{bdt_color}}
\end{figure}

To validate that the BDT has learned physically meaningful charge information, we discretize its output to the nearest integer and color-code events in the original signal space. Fig.~\ref{bdt_color}\,(a) shows seed versus second-largest channel values, where different charges form clearly separated bands. In the high-amplitude region, seed saturation causes high-$Z$ events to overlap along the seed axis. Fig.~\ref{bdt_color}\,(b) shows second- versus third-largest channel values; in the saturated region, the secondary channels still provide distinguishable band structures, confirming that the BDT effectively combines multi-channel information across different signal regimes.

\begin{figure}[htbp]
\centering
\includegraphics[width=0.45\hsize]{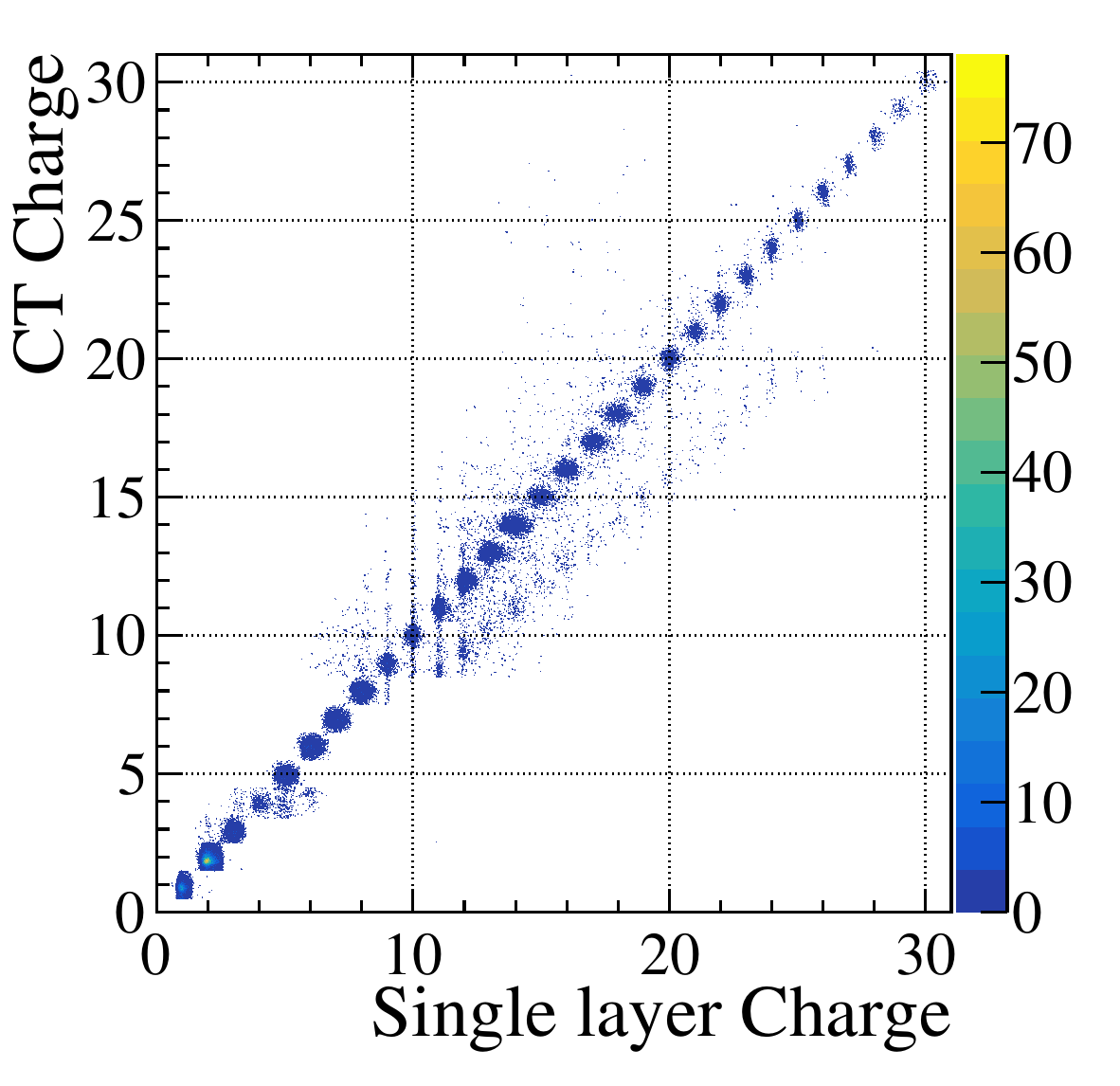}
\caption{Two-dimensional comparison between BDT single-layer charge output and CT charge on the testing dataset. The diagonal structure confirms agreement between the two independent measurements.\label{testing_z_vs_ct}}
\end{figure}

We further compare the BDT charge with the CT measurement on the test dataset, as shown in Fig.~\ref{testing_z_vs_ct}. The two independent measurements show overall consistency along the diagonal, confirming that the BDT has learned a physically meaningful charge mapping.

\section{Track reconstruction for nuclei}

In addition to charge measurement, the telescope provides precise position measurements, which enable studies of the spatial resolution and the performance of different regions of a detector under test.

\subsection{Track finding with PID information}

The heavy-ion beam used in this study was a secondary mixed beam produced by fragmentation, and a single triggered event often contains multiple particles. The cluster multiplicity per detector layer increases significantly with the charge of the highest-$Z$ nucleus in the event: for helium ($Z=2$) the typical multiplicity is about 1, for carbon ($Z=6$) it rises to 2--3, and for heavy nuclei such as iron ($Z=26$) it reaches approximately 40. This high cluster multiplicity makes track reconstruction in mixed beams challenging. In principle, one could enumerate all possible combinations of one cluster per detector layer and select the candidate with the best fit quality, for example, the one with the minimum $\chi^2$. In practice, however, this quickly becomes computationally uneconomical when the cluster multiplicity is high. Moreover, in this study, our purpose is not merely to find the best-fitted trajectory, but to identify the track corresponding to the particle with the highest charge in the event, since heavy nuclei are relatively rare in the mixed beam. To address both issues, we developed a track-finding algorithm that explicitly incorporates PID information. The event-by-event procedure is as follows.

\begin{figure}[h]
\centering
\includegraphics[width=0.9\hsize]{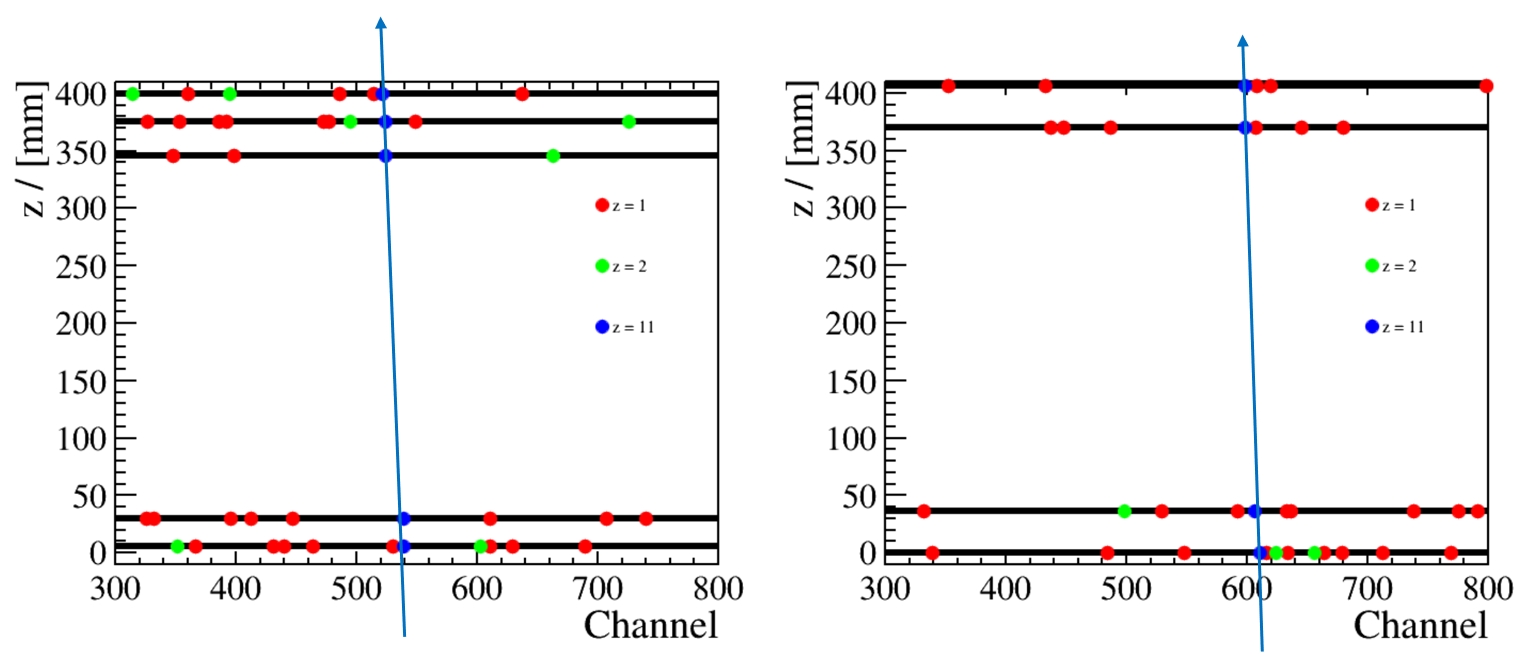}\\
\caption{Event display of a typical event. The black solid lines represent individual detector layers, with the left panel showing the 5 $x$-layers and the right panel showing the 4 $y$-layers. Different colors indicate clusters with different charge labels assigned by the BDT. The blue arrow denotes the reconstructed primary track.} \label{event_display}
\end{figure}

\begin{itemize}
    \item For each layer, assign charge labels to all available clusters using the trained BDT;
    \item Select the cluster with the highest charge across all layers and record the rounded maximum charge value, $Z_{\max}$;
    \item For each layer, identify all clusters with charge values in the range $[Z_{\max}-1, Z_{\max}]$;
    \item If every layer contains at least one valid cluster, construct the primary track by selecting the highest-quality trajectory.
\end{itemize}

Fig.~\ref{event_display} shows a typical event display in which the primary track is successfully reconstructed. For heavy nuclei, associated lighter fragments are often present. In such events, utilizing PID information significantly reduces the computational load and increases the probability of selecting the track corresponding to the heaviest nucleus.

\subsection{Incident position reconstruction with eta algorithm\label{pca}}

\begin{figure}[h]
\centering
\subfloat[$Z=2$]{\includegraphics[width=0.32\textwidth]{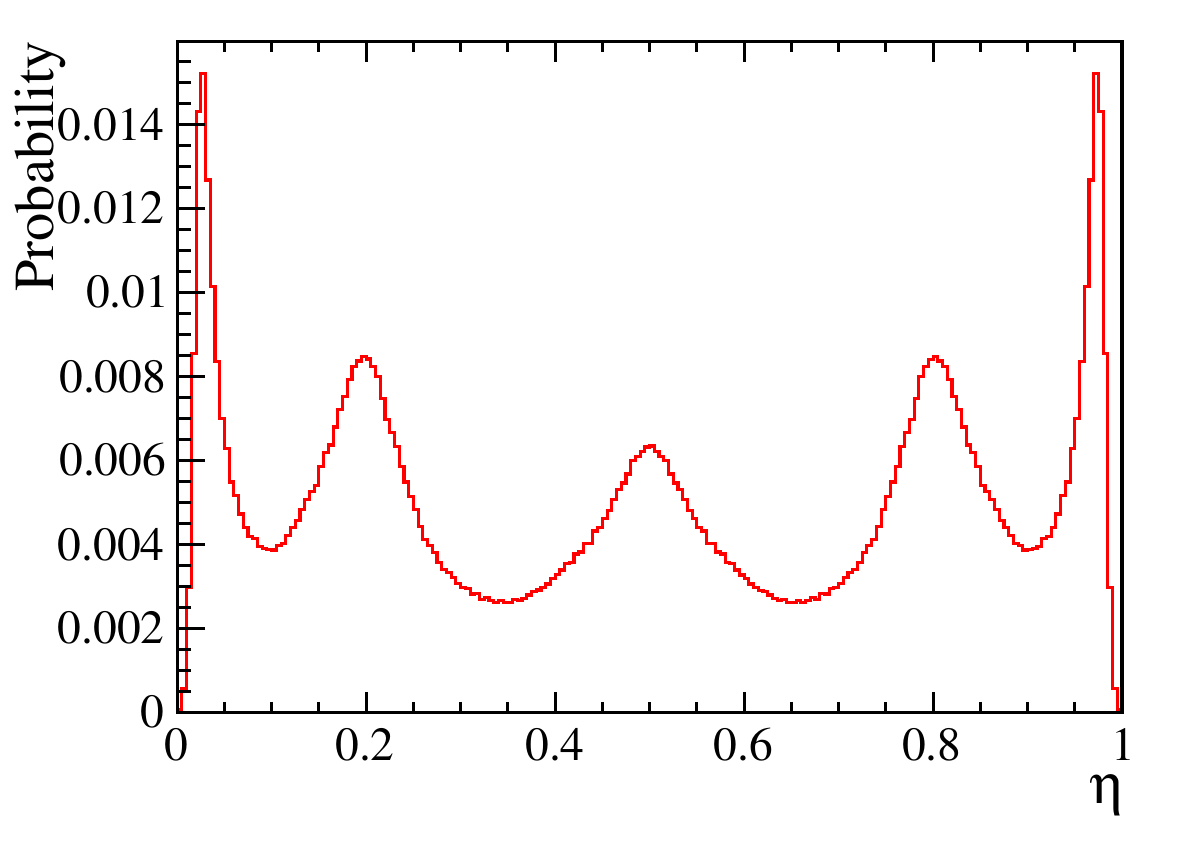}}
\subfloat[$Z=6$]{\includegraphics[width=0.32\textwidth]{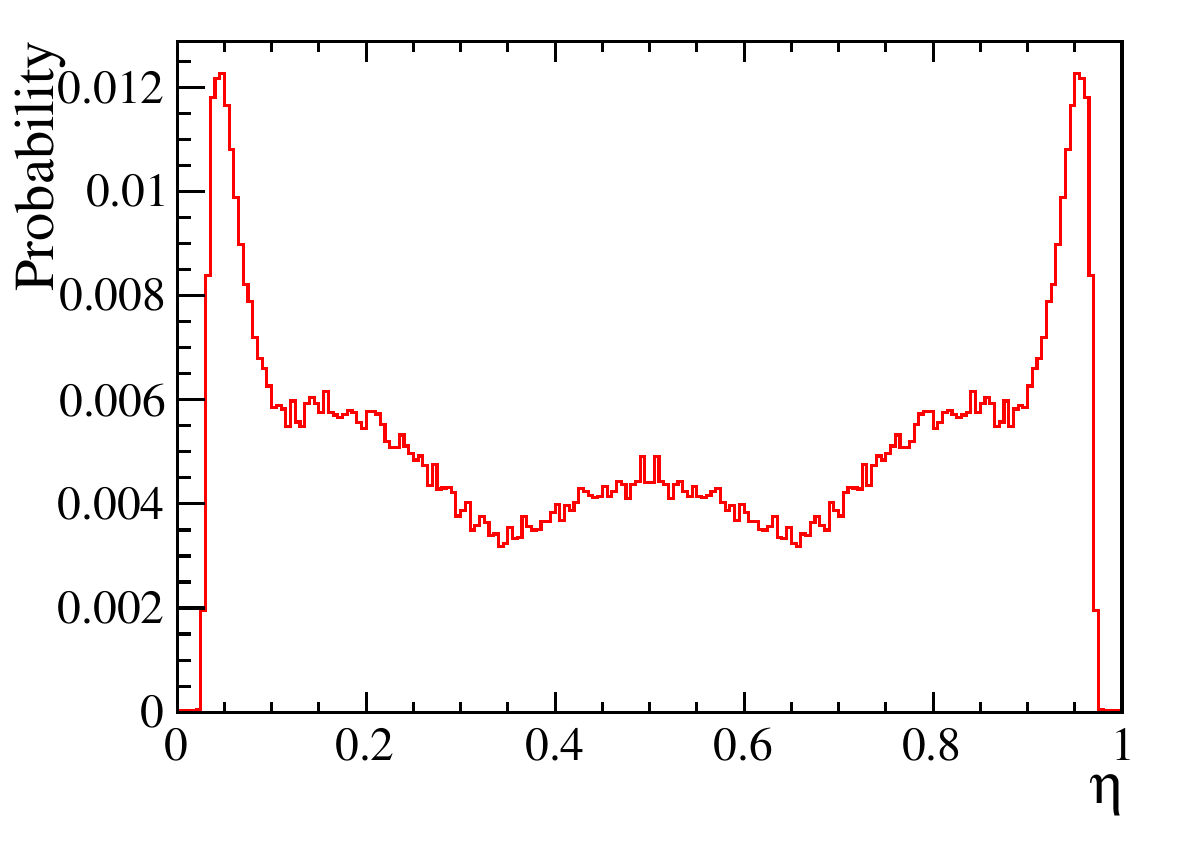}}
\subfloat[$Z=10$]{\includegraphics[width=0.32\textwidth]{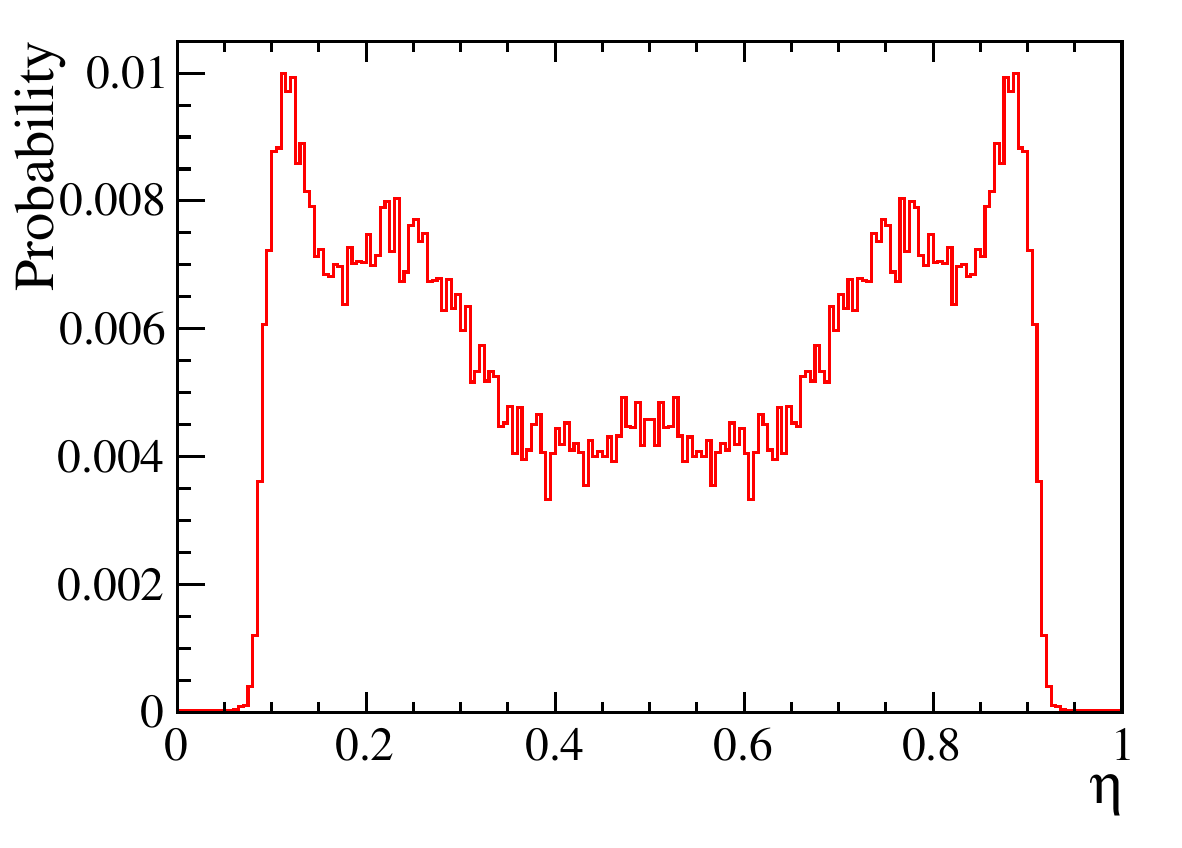}} \\
\subfloat[$Z=14$]{\includegraphics[width=0.32\textwidth]{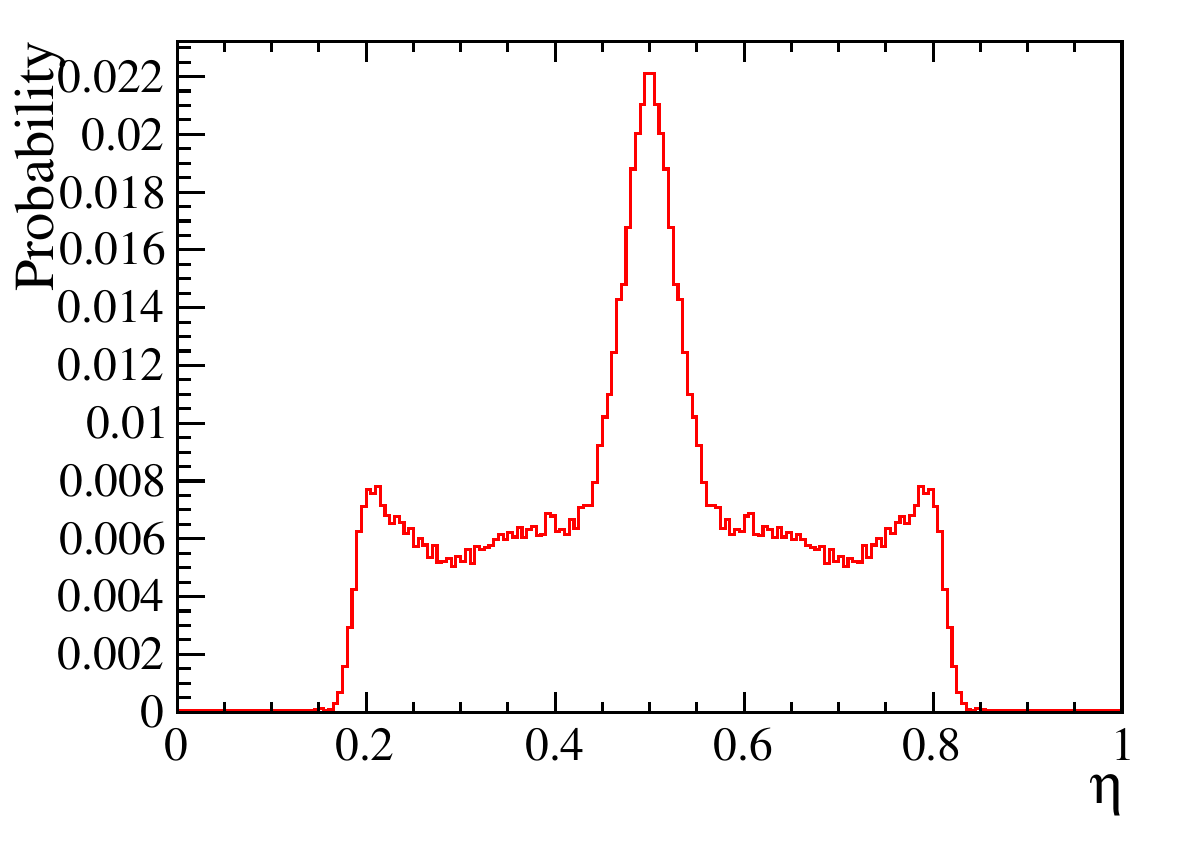}}
\subfloat[$Z=20$]{\includegraphics[width=0.32\textwidth]{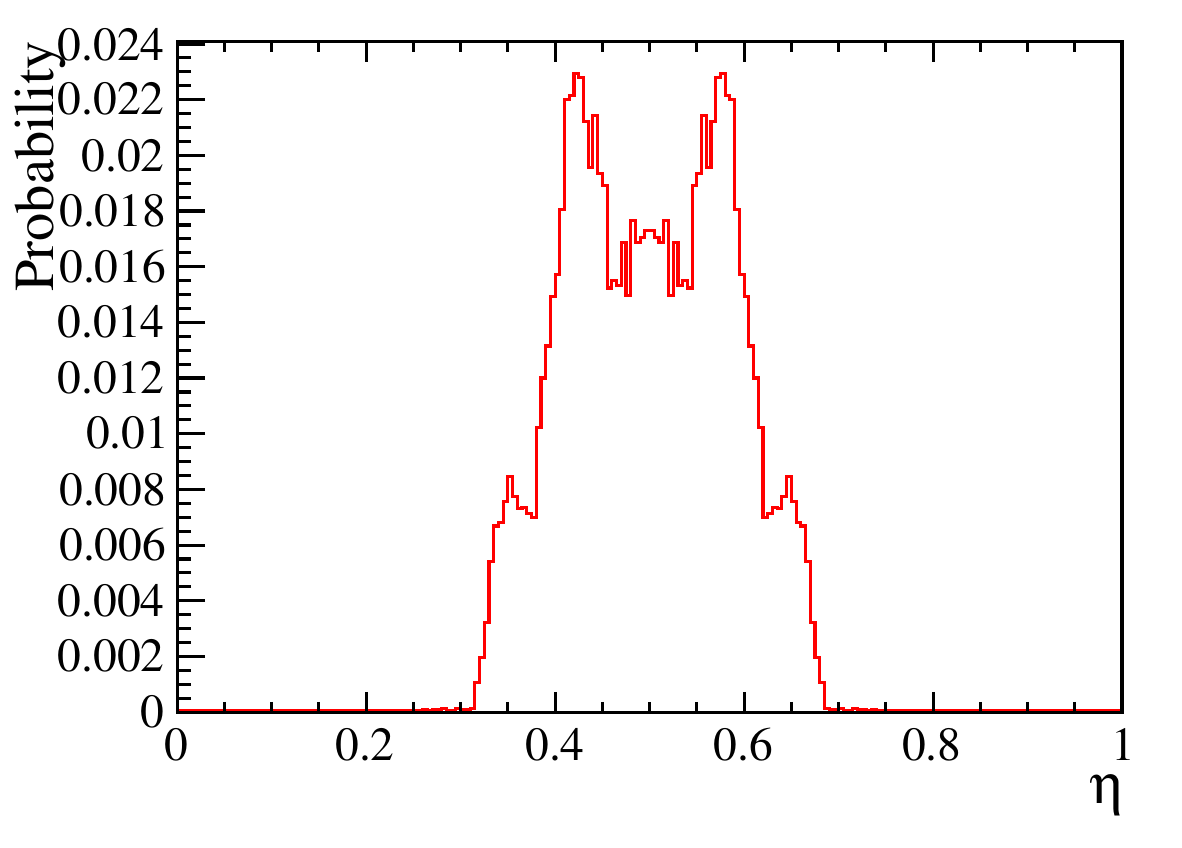}}
\subfloat[$Z=26$]{\includegraphics[width=0.32\textwidth]{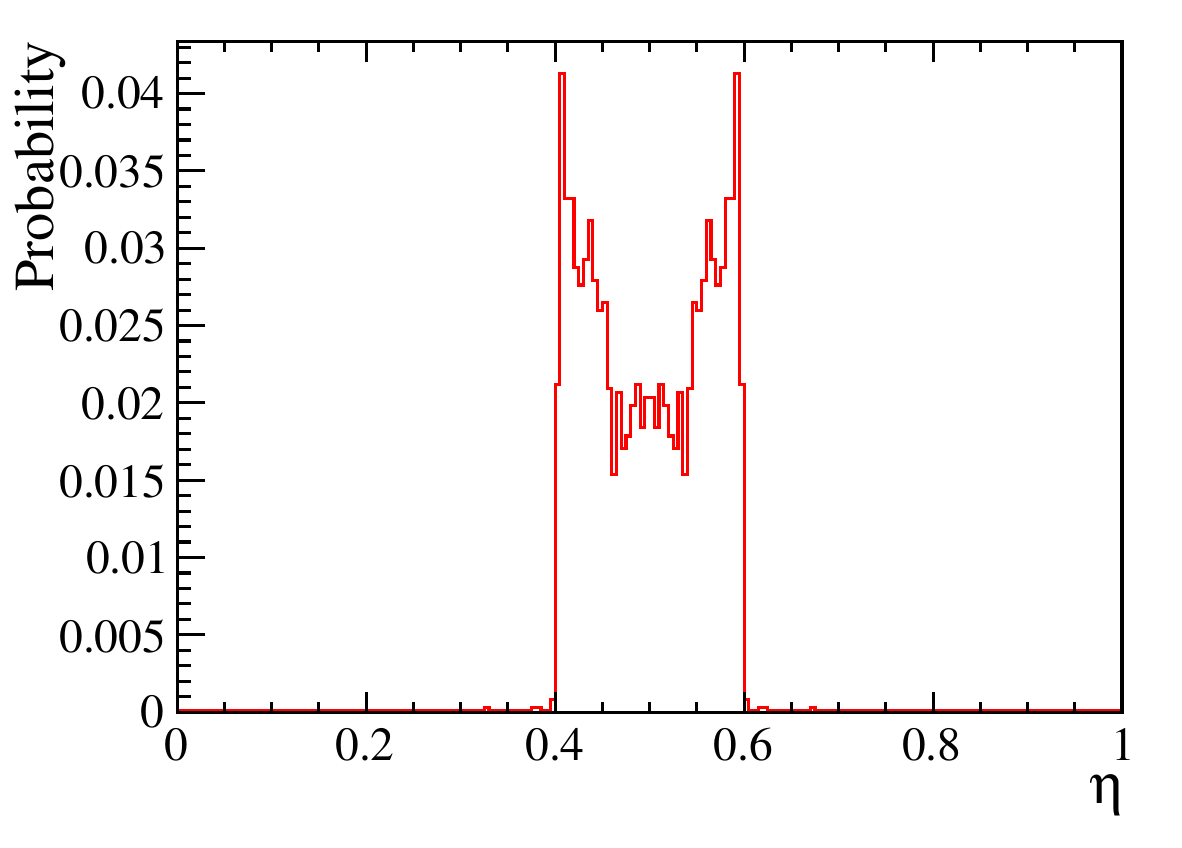}}
\caption{$\eta$ distribution of different nuclei.} \label{eta_dist}
\end{figure}

Many studies have shown that for silicon microstrip detectors with normally incident particles, the $\eta$ algorithm is the optimal position reconstruction method~\cite{Belau:1983eh,Turchetta:1993vu}. We therefore adopt this algorithm for the spatial resolution study with heavy nuclei. Using the variable $\eta$ defined in the previous section, the incident position $x$ is given by a nonlinear function of $\eta$:
\begin{equation}
x_{\eta} = x_L + Pf(\eta) \;,
\end{equation}
where $P$ is the readout strip pitch and $f(\eta)$ is a monotonically increasing function of $\eta$.

For the case of normal incidence, assuming a uniform distribution of particles hitting between any two adjacent readout strips, $f(\eta)$ is given by~\cite{Turchetta:1993vu}:
\begin{equation}
    f(\eta_0) = \frac{\int_{0}^{\eta_0}{\frac{dN}{d\eta} d\eta}}{\int_{0}^{1}{\frac{dN}{d\eta} d\eta}} = \int_{0}^{\eta_0}{\phi(\eta) d\eta} \;,
\end{equation}

where $\phi(\eta)$ is the probability density function (PDF) of the $\eta$ distribution. Fig.~\ref{eta_dist} shows the $\eta$ distributions obtained for several different nuclei. For light nuclei ($Z=1,\, 2, \,3$), five peaks are observed, corresponding to the two readout strips and the three floating strips between them. For heavier nuclei, the larger energy deposition leads to a wider electron--hole cloud, and the central peaks become less pronounced. For nuclei with $Z \geq 15$, due to the large signals and extensive charge sharing among neighboring channels, there are no events near $\eta = 0$ or $1$, and most events shift toward $\eta = 0.5$. Using these $\eta$ distributions, we can accurately reconstruct the impact position of each incident particle.

\subsection{Alignment and track fitting}

The track fitting is performed using the General Broken Lines (GBL) algorithm~\cite{Kleinwort:2012np}, which is used in experiments such as AMS-02~\cite{Yan:2023xtc} and CMS~\cite{Otarid:2023anx}. GBL is an advanced track-fitting method that accounts for multiple scattering effects at each detector layer. An additional advantage is that it provides the complete covariance matrix of all track parameters, making it well-suited as a track model for calibration and alignment with Millepede~II~\cite{Blobel:2011az}, a global parameter optimization algorithm capable of handling up to hundreds of thousands of parameters. To minimize systematic errors and independently verify the results, we use helium data for alignment and subsequently obtain spatial resolution results for all other nuclei.

\section{Performance \label{sec:performance}}

\subsection{Charge resolution of the telescope}

Using the BDT trained with the method described in Sec.~\ref{sec:training_set}, we performed charge identification for a total of $5\, \mathrm{M}$ events. Since each detector layer provides an independent charge measurement, for reconstructed tracks as described in Sec.~\ref{pca}, a mean charge value is calculated across all 9 layers, representing the overall charge measurement by the telescope. To mitigate the influence of the Landau tail, we employ a ``truncated'' averaging method by discarding the highest charge values before computing the mean from the remaining measurements. 

\begin{figure*}[htbp] \centering 
\subfloat[\label{}]{\includegraphics[width=0.8\hsize]{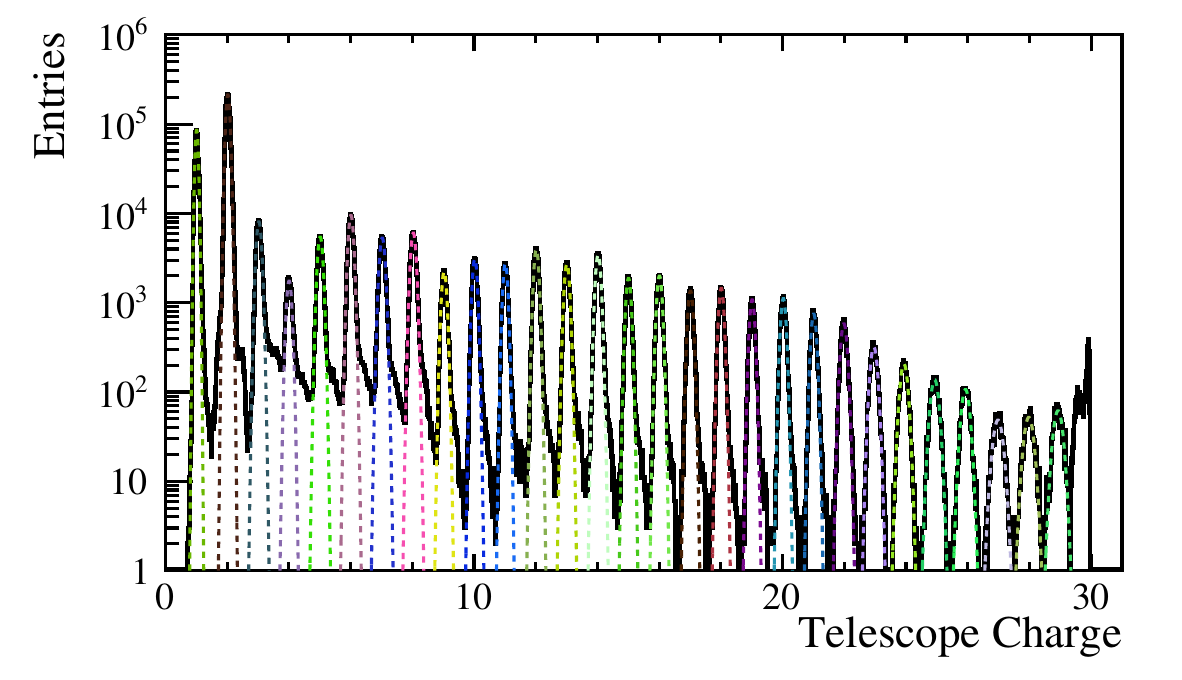}}
\\
\subfloat[\label{}]{\includegraphics[width=0.8\hsize]{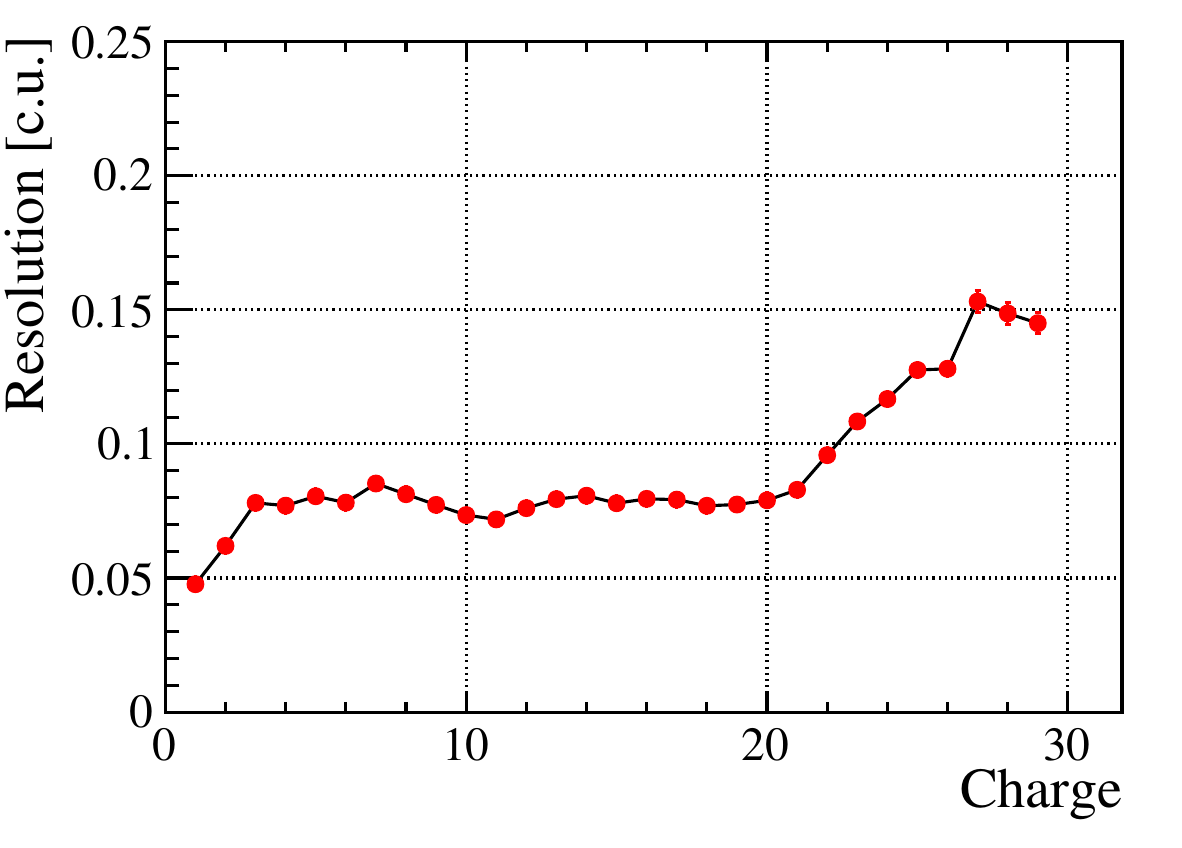}}
\caption{(a) Mean charge distribution of the 9-layer telescope for $5\, \mathrm{M}$ events. Each peak is fitted with a Gaussian function, and the element symbols are labeled. (b) Charge resolution (Gaussian sigma) as a function of nuclei charge $Z$ for the telescope.}\label{charge:all}
\end{figure*}

The resulting mean charge distribution is shown in Fig.~\ref{charge:all} (a), where each peak is fitted with a Gaussian function. The sigma of the Gaussian fit represents the charge resolution for the corresponding nucleus. The charge resolution of the telescope is summarized in Fig.~\ref{charge:all} (b). For nuclei from \( Z = 4\, \mathrm{(Be)} \) to \( Z = 20 \, \mathrm{(Ca)} \), the charge resolution is around 0.08 charge units. For heavier nuclei up to \( Z = 29 \, \mathrm{(Cu)} \), the resolution gradually degrades to 0.16 charge units. The reason for this is that the seed values of an increasing fraction of events reach saturation, see Fig.~\ref{channel_corr} (a).

\subsection{Spatial resolution of a single layer}

To characterize the intrinsic spatial resolution of a single detector layer at a beam momentum of $\sim 150\,\mathrm{GeV/n}$, we select the middle layer as the Device Under Test (DUT), which is excluded from the track fitting. A track is fitted using the remaining eight reference layers with the GBL algorithm and extrapolated to predict the incident position on the DUT plane. The difference between the predicted position and the position reconstructed by the DUT forms the unbiased residual.

\begin{figure*}[htbp] \centering
\subfloat[$Z=1$ (H)\label{res:z1}]{\includegraphics[width=0.32\hsize]{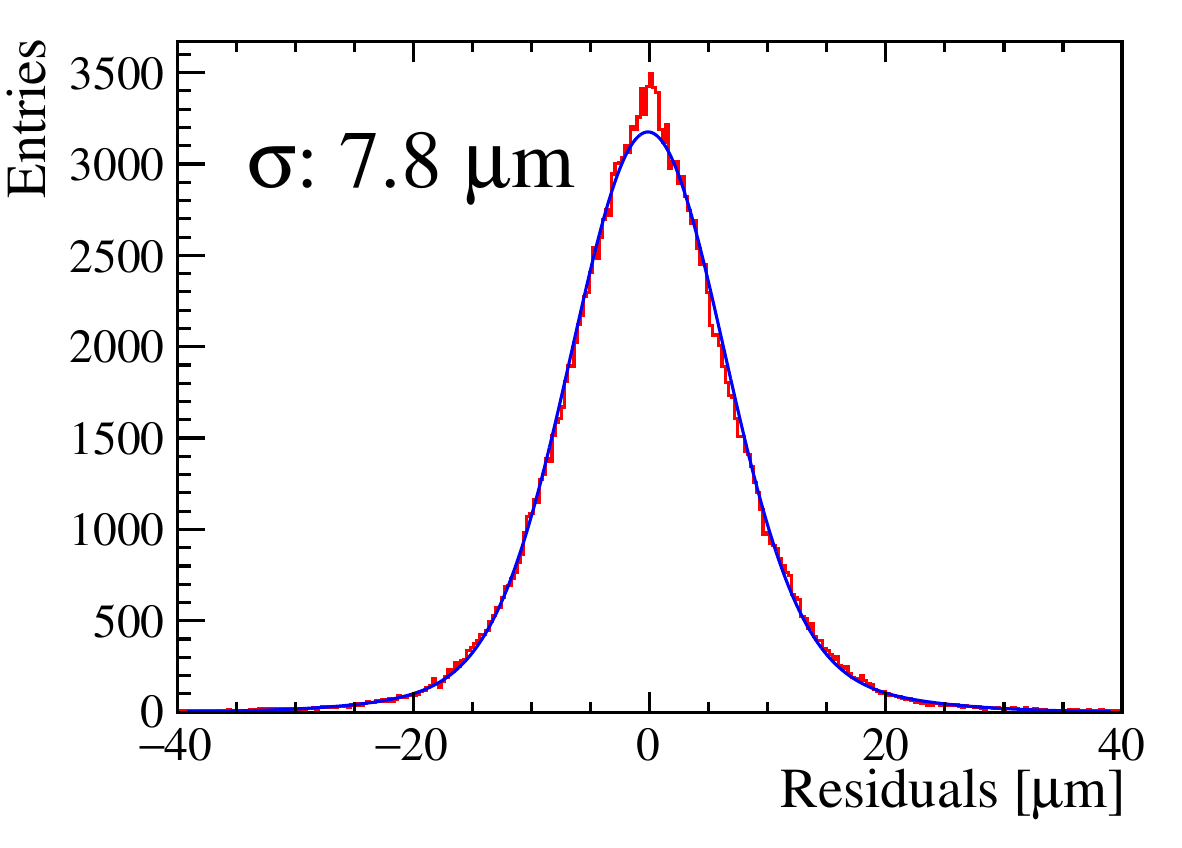}}
\hfill
\subfloat[$Z=2$ (He)\label{res:z2}]{\includegraphics[width=0.32\hsize]{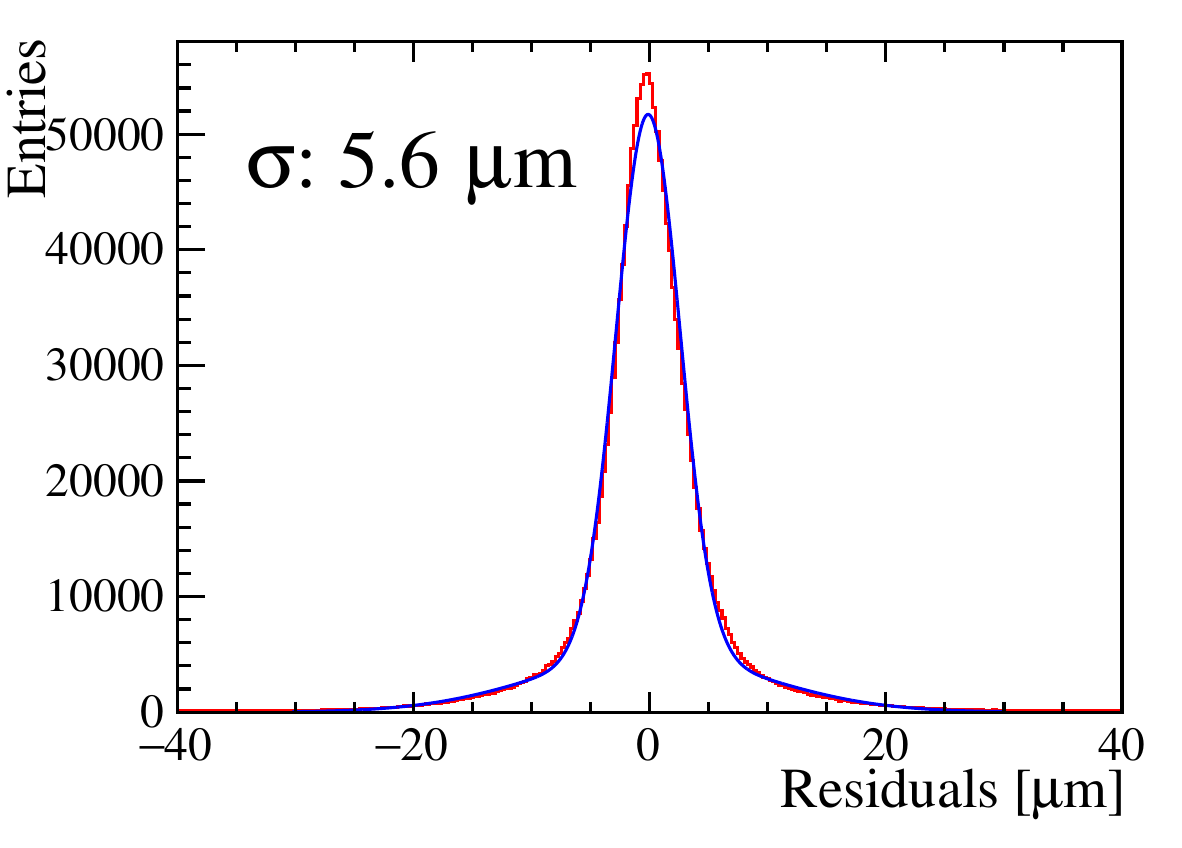}}
\hfill
\subfloat[$Z=6$ (C)\label{res:z6}]{\includegraphics[width=0.32\hsize]{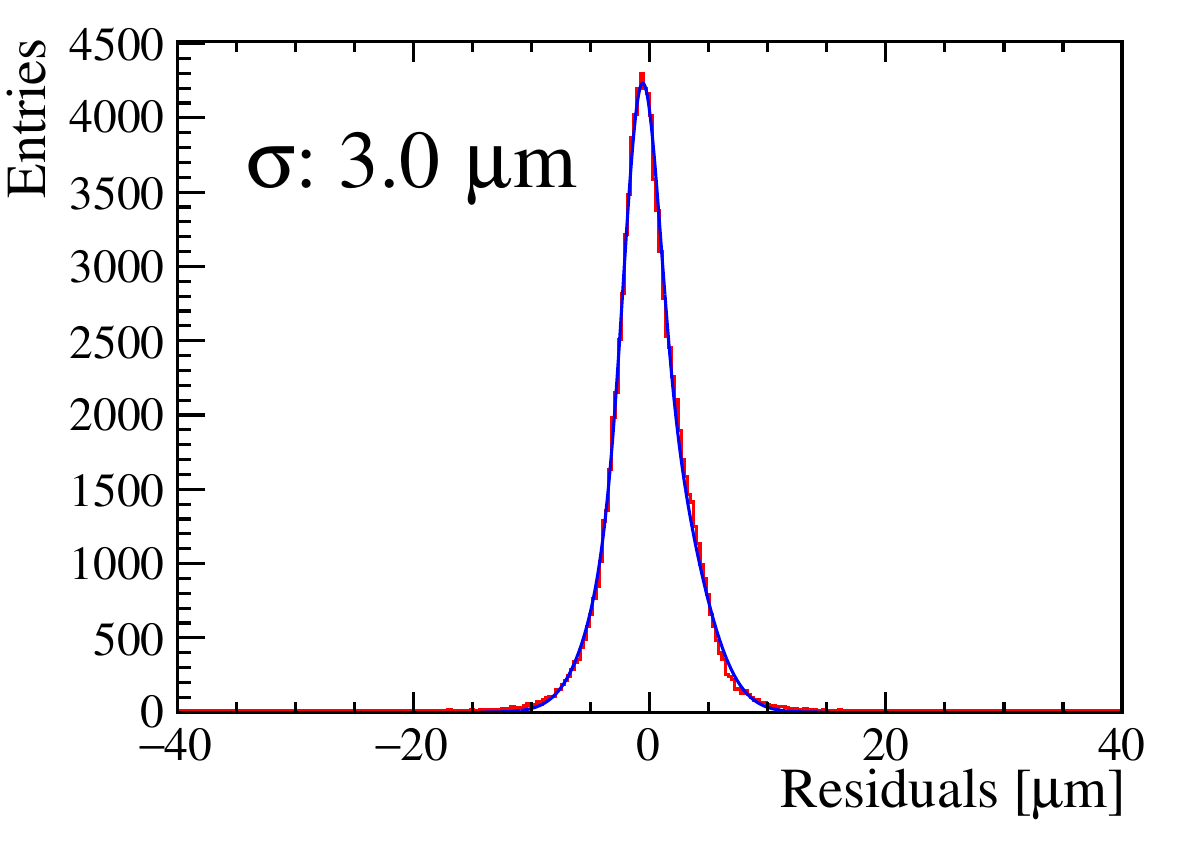}}
\\
\subfloat[$Z=14$ (Si)\label{res:z14}]{\includegraphics[width=0.32\hsize]{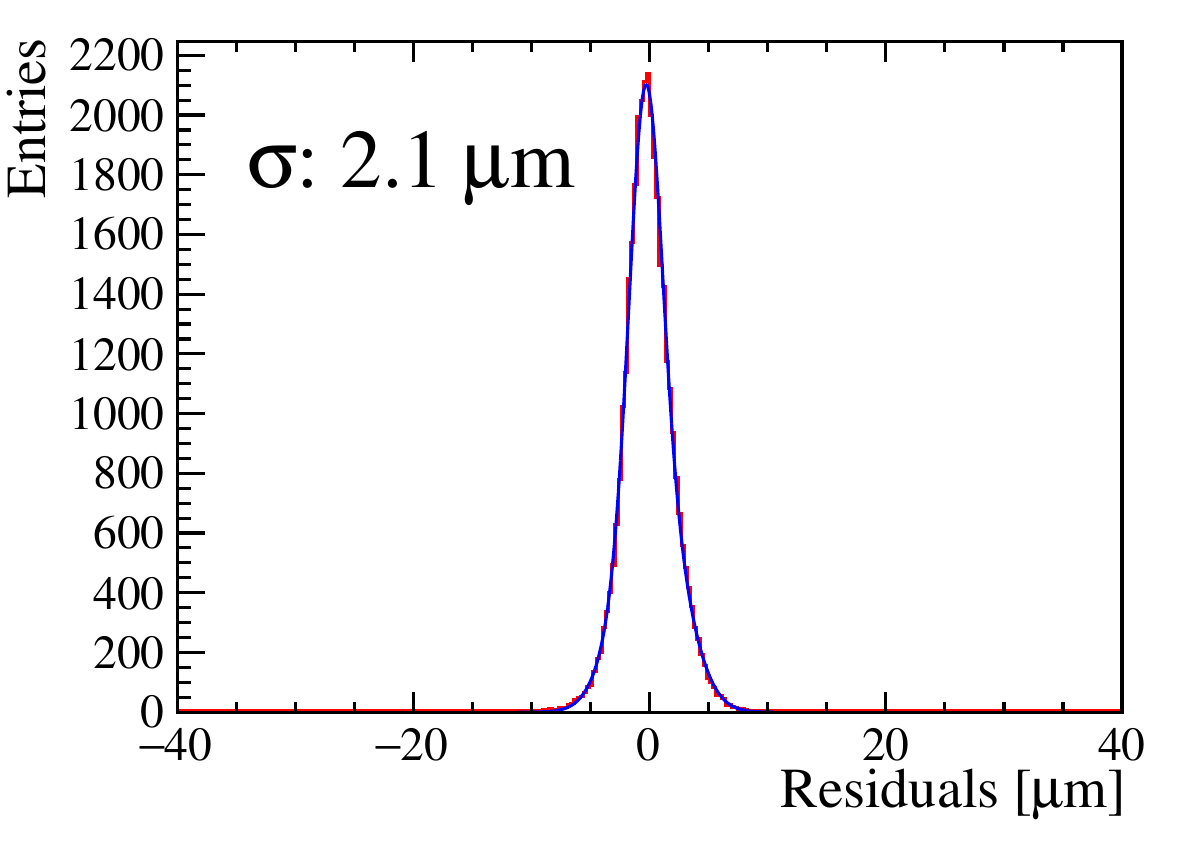}}
\hfill
\subfloat[$Z=20$ (Ca)\label{res:z20}]{\includegraphics[width=0.32\hsize]{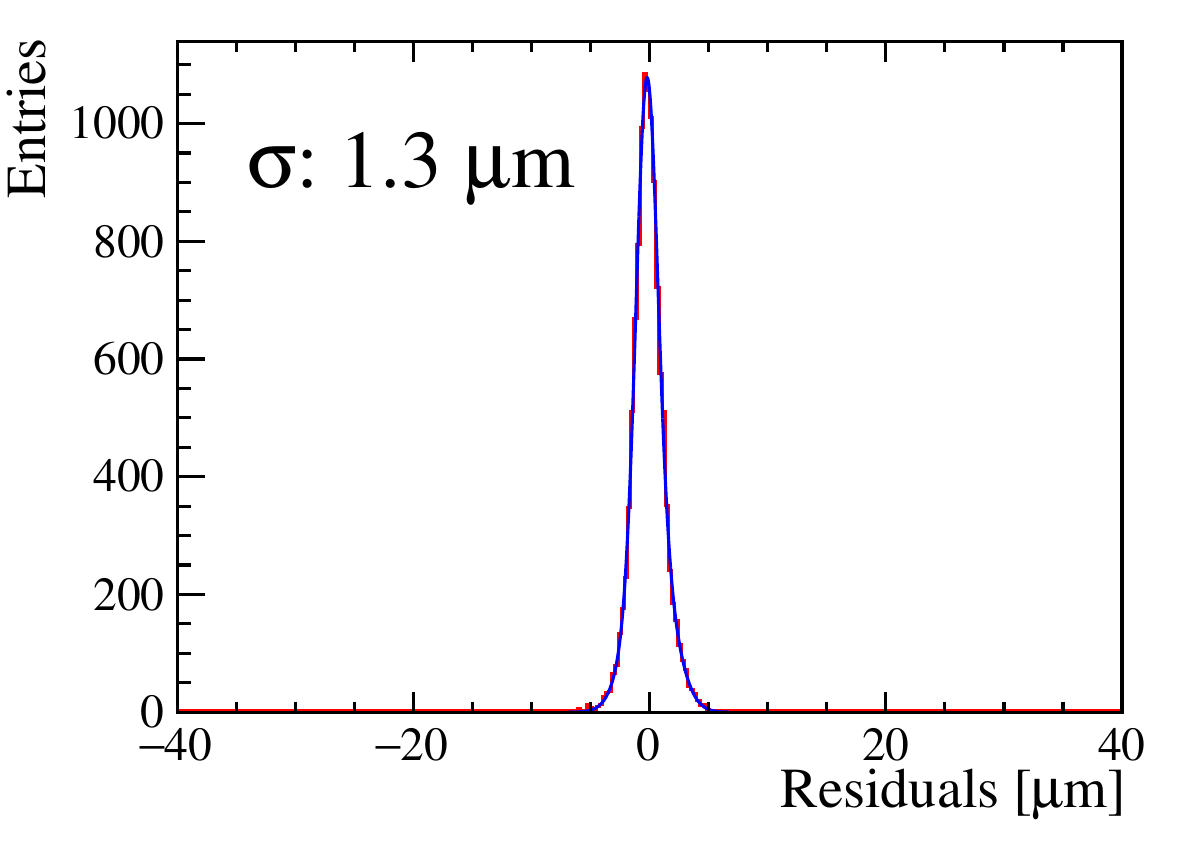}}
\hfill
\subfloat[$Z=26$ (Fe)\label{res:z26}]{\includegraphics[width=0.32\hsize]{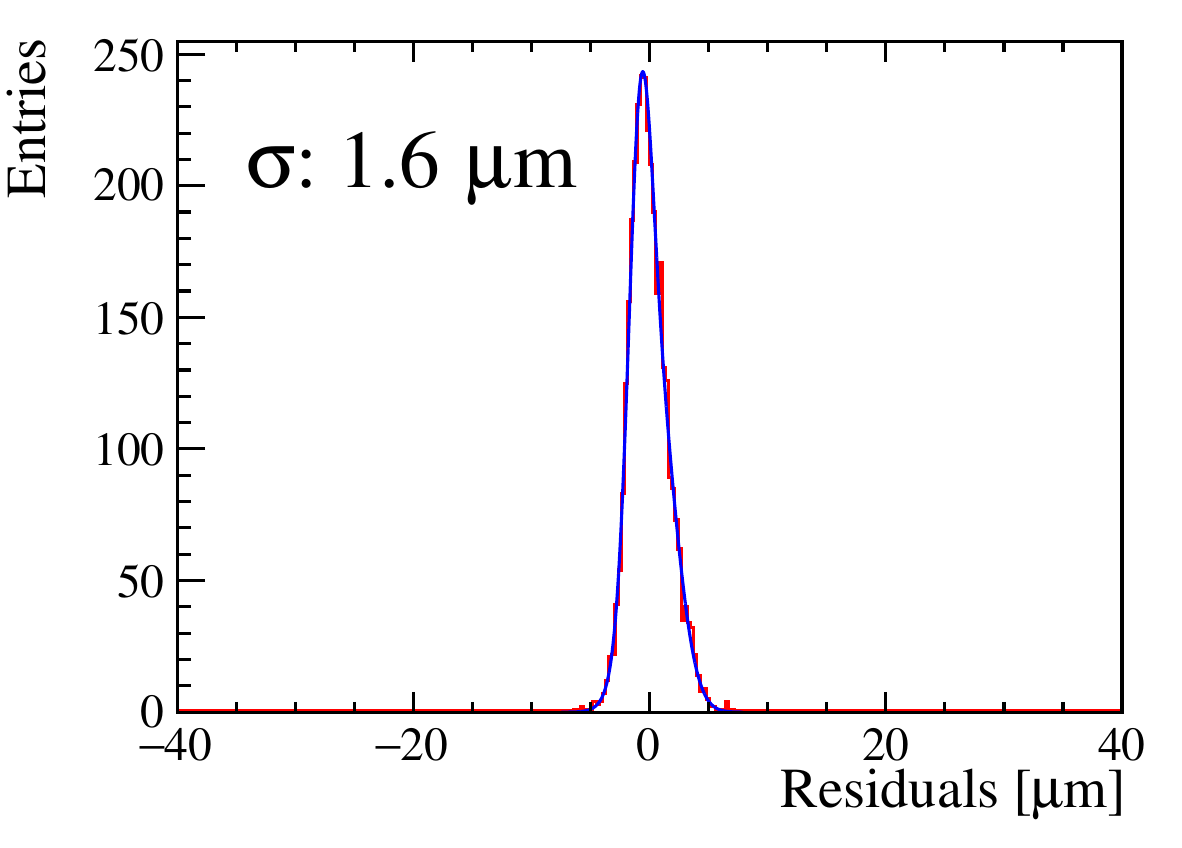}}
\caption{Unbiased residual distributions of the DUT for six representative nuclei. Each distribution is fitted with a double Gaussian function.}\label{residual_dist}
\end{figure*}

\begin{figure}[htbp]
\centering
\includegraphics[width=0.8\hsize]{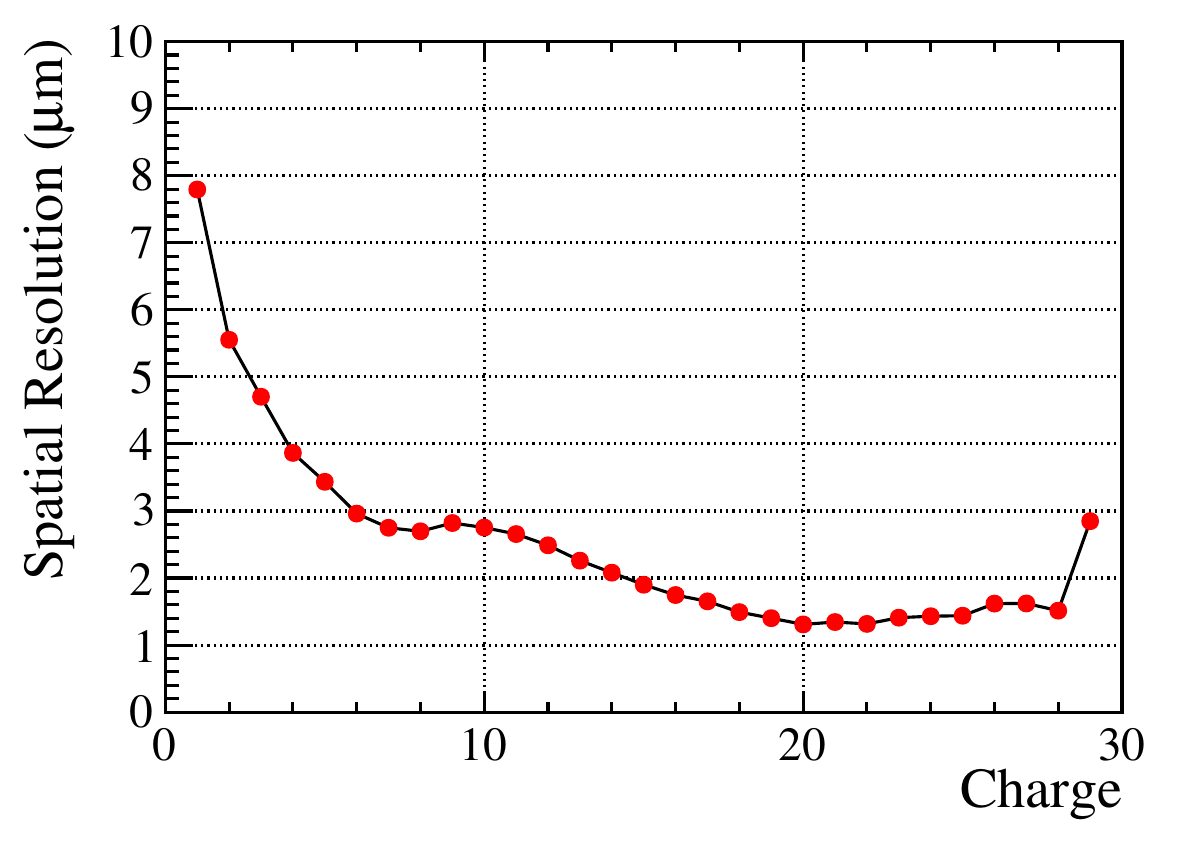}
\caption{Spatial resolution of a single layer as a function of nuclei charge $Z$ at a beam momentum of $\sim 150\,\mathrm{GeV/n}$.} \label{res_dist}
\end{figure}

Fig.~\ref{residual_dist} shows the unbiased residual distributions for six representative nuclei spanning the full charge range. Each distribution is fitted with a double Gaussian function, and the weighted sigma is taken as the spatial resolution. The spatial resolution of the DUT as a function of $Z$ is presented in Fig.~\ref{res_dist}. For protons ($Z=1$), the resolution is approximately \SI{7.8}{\micro\metre}. As $Z$ increases, the signal amplitude grows, leading to a higher signal-to-noise ratio and improved position reconstruction. The resolution reaches approximately \SI{3.0}{\micro\metre} for carbon ($Z=6$) and a minimum of approximately \SI{1.5}{\micro\metre} around $Z = 20\text{--}22$. For nuclei with $Z \geq 26$, the spatial resolution shows some degradation, which could be attributed to the saturation of the seed channel. In this case, the $\eta$ value is mainly determined by variations of the second-largest channel, since the seed channel remains fixed at the saturation value. As a result, the sensitivity of $\eta$ to the particle impact position is reduced, leading to a deterioration of the position resolution.

\subsection{Telescope pointing resolution}

The intrinsic spatial resolution of a single detector layer has been discussed in the previous section. When multiple layers are combined to form a tracking telescope, the accuracy with which the particle trajectory can be predicted at an arbitrary position along the beam line depends not only on the single-layer resolution but also on the geometric configuration of the telescope and the beam conditions.

The telescope pointing resolution $\sigma_{\mathrm{tele}}$ is mainly determined by two contributions. The first is a geometric term arising from the intrinsic spatial resolution of the telescope layers and their relative positions along the beam direction. For straight-line tracks, this contribution can be estimated analytically from the propagation of the track-fit covariance matrix obtained from a linear fit to the measured hit positions~\cite{Jansen:2016bkd}. The second contribution originates from multiple Coulomb scattering in the detector material, which depends on the beam energy, the charge of the incident particle, and the material budget of the telescope planes~\cite{ParticleDataGroup:2024cfk}.

Both effects are naturally incorporated in the General Broken Lines (GBL) track fitting framework, which accounts for measurement uncertainties and scattering processes simultaneously and provides the optimal estimate of the track parameters and their covariance matrix~\cite{Kleinwort:2012np}. The resulting covariance matrix can be propagated to any position along the beam axis to obtain the corresponding telescope pointing resolution.

For a given telescope geometry and beam configuration, the pointing resolution can also be evaluated using dedicated Monte Carlo simulations that include the detector resolution and material distribution. Such simulations provide a practical way to quantify the expected tracking performance of the telescope under realistic experimental conditions.

\section{Summary and prospect}

We have designed a beam telescope system based on silicon microstrip detectors for charge and position measurements in heavy ion beam tests. The system was thoroughly characterized during a heavy ion beam campaign conducted at CERN SPS in November 2023. To address the limitation that charge measurements with silicon strip detectors strongly rely on prior information, we developed a heavy nuclei charge measurement algorithm based on BDT and hybrid machine learning techniques. With this approach, charge measurement was achieved on a large dataset ($5\, \mathrm{M}$ events) using a training sample of $0.4\, \mathrm{M}$ events labeled by a small external charge tagger. The 9-layer telescope achieves a charge resolution better than 0.11 charge units for nuclei from $Z = 1$ to $Z = 22$, and better than 0.16 charge units for nuclei up to $Z = 29$, enabling effective charge separation across a wide range of species. Moreover, by exploiting the independent charge measurements provided by multiple detector layers, we developed a track-finding algorithm incorporating PID information, which significantly reduces computational complexity. The single-layer spatial resolution reaches approximately \SI{1.5}{\micro\metre} for nuclei around $Z = 20\text{--}22$ and is approximately \SI{7.8}{\micro\metre} for protons ($Z = 1$).

Since charge measurement is independently performed at each detector layer, increasing the number of layers could enhance the overall charge resolution. In fact, the telescope has been expanded from 9 to 12 layers in the latest configuration. Furthermore, based on previous studies, adjusting the charge amplitude response configuration of the IDE1140 chip and employing thinner SSDs could extend the system’s capability to identify heavier nuclei. In the present application of this beam telescope, only normal incidence needs to be considered. For practical experimental conditions involving various incident angles, incorporating angle information as an additional input feature for BDT training would be necessary. The hybrid machine learning algorithm, requiring only a small number of labeled samples, could serve as an effective charge reconstruction method for spaceborne cosmic ray experiments such as AMS-02, DAMPE, and HERD.

\section*{Acknowledgments}
This study was supported by the National Key Programme for S\&T Research and Development (Grant NO.: 2022YFA1604800), and the National Natural Science Foundation of China (Grant NO.: 12342503). We express our gratitude to our colleagues in the CERN accelerator departments for the excellent performance of the SPS. We also thank our colleagues at Fudan University and INFN, Florence, for supplying the scintillator and CT detectors used in this work. We appreciate helpful discussions with Shudong Wang (IHEP) and Jiaoyang Xu (BNU).

\bibliography{reference}
\bibliographystyle{elsarticle-num}

\end{document}